\documentclass[11pt,a4paper]{article}
\usepackage{packages}
\usepackage{jheppub}

\numberwithin{equation}{section}

\title{Mapping Dirac gaugino masses}

\author{Steven Abel}
\author{and Daniel Busbridge}

\affiliation{Institute for Particle Physics Phenomenology, \\ Durham University, South Road, Durham, DH1 3LE, UK}

\emailAdd{s.a.abel@durham.ac.uk}
\emailAdd{d.w.busbridge@durham.ac.uk}

\abstract{\small \noindent We investigate the mapping of Dirac gaugino masses through regions of strong coupling, focussing on
SQCD with an adjoint. These models have a well-known Kutasov duality, 
under which a weakly coupled electric UV description can flow to a different weakly coupled magnetic IR description. We provide evidence to show that 
Dirac gaugino mass terms map as 
\[
\lim_{\mu\rightarrow\infty}  
\frac{m_{D}}{g \kappa^{\frac{1}{k+1}}} = 
\lim_{\mu\rightarrow 0}  \frac{\tilde m_{{D}}}{\tilde{g} \tilde{\kappa}^{\frac{1}{k+1}}}
\]
under such a flow, where 
the coupling $\kappa$ appears in the superpotential of the canonically normalised theory as $W\supset \kappa X^{k+1}$. This combination is an RG-invariant to all orders in perturbation theory, but establishing the mapping in its entirety is not straightforward 
because Dirac masses are not the spurions of holomorphic couplings in the 
$\NN=1$ theory. To circumvent this, we first present evidence that deforming the Kutasov theory can make it 
flow to an $\NN=2$ theory with parametrically small $\NN=1$ deformations. This is shown to happen perturbatively in the weakly coupled theory, and we also identify the higgsing mechanism that has to take place in the dual theory. This is seen to occur correctly even when both theories are at strong coupling. Using harmonic superspace 
techniques we then identify the prepotential that can induce the same $\NN=1$ deformations in the presence of electric and magnetic FI-terms. We show that the correct $\NN=1$ scalar potential and fermion lagrangian are generated.
It is then shown that pure Dirac mass terms can be induced by the same mechanism, and we find that the proposed RG-invariant is indeed preserved under $\NN=2$ duality, and thence along the flow to the dual $\NN=1$ Kutasov theories. Possible phenomenological applications are discussed.}


\arxivnumber{1306.6323}

\begin{document}
\maketitle

\newpage

\section{Introduction and Summary}

There is continuing interest in the role that Dirac gauginos may play in supersymmetry due to their possible physical advantages
over the Majorana variety~\cite{Fayet1978, Polchinski1982, Hall1991, Fox2002, Nelson2002a,Antoniadis2005,Antoniadis2006,Antoniadis2006a,Hsieh2008,Amigo2008,Choi2008a,Choi2008,Blechman2009,Benakli2008, Belanger2009,Choi2009,Benakli2010a, Chun2010, Benakli2010,Carpenter2010,Kribs2010,Choi2010,Abel2011a,Davies2011a,Benakli2011,Benakli2011a,Heikinheimo2011,Itoyama2011,Kribs2012,Davies2012,Goodsell2012,Benakli2012,Itoyama2013}. Most of this work however considers Dirac gaugino masses in a perturbative setting: supersymmetry is broken at some high scale, and this leads to mass terms that can be calculated within the perturbation theory of the low energy effective theory. This is true even if, as in \cite{Abel2011a}, the adjoint fermions that partner the gauginos in the Dirac mass term  (part of the so-called ESP supermultiplet) are themselves the mesinos of some strongly coupled ${\cal N}=1$ gauge theory. Likewise, within supersymmetric Randall-Sundrum set-ups, Dirac mass terms can appear when the gaugino zero-modes of bulk gauge fields marry with the lowest lying KK modes \cite{Marti2001,Abel2010a,Argurio2012}. In either case the gauge symmetry of interest is just a flavour symmetry of the strongly coupled physics. Moreover if Dirac mass terms do originate from operators of an ultra-violet (UV) theory that becomes strongly coupled, they are trivial to map to the infra-red (IR) due to holomorphy, being given simply by a combination of SUSY breaking, transmutation and fundamental scales.  For example in SQCD the adjoint field $\psi_X$ can be a mesino mapped (upto an unknowable normalization factor) as $\psi_X\sim \Lambda^{-1} \tilde{Q}.\psi_Q$, where $Q$ indicate (s)quarks of the confining UV theory. Then the effective Dirac mass term coupling this state to a flavour gaugino would arise from the non-renormalizable operator 
\begin{equation}
W \supset \frac{\tilde{Q}.Q \,\WW^\alpha \WW^\prime_\alpha}{M^2} , 
\label{diracterm} 
\end{equation}
 where $\langle \WW^\prime_\alpha \rangle \sim \theta_\alpha \,D $ is a supersymmetry breaking spurion $D$ term, and $M$ is some fundamental scale. Obviously $m_D\sim \Lambda D/M^2$ is about all one can say in this case. 

A more interesting question is what happens to Dirac mass terms involving the gauginos of the colour gauge symmetry that becomes strongly coupled. Can such terms be mapped from UV to IR and if so how do they appear in the IR physics? Conversely, can Dirac mass terms in the IR be mapped from operators in the UV? To make the question precise, we will focus on the ${\cal N}=1$ adjoint+QCD duality of ref. \cite{Kutasov1995} which we refer to as SQCD+$X$ (occasionally as Kutasov) duality. 
These models and many variants were analysed 
in refs.\cite{Kutasov1996,Brodie1998,Intriligator1995b}, and phenomenological applications have been suggested in many works. For our current purposes SQCD+$X$ is precisely the context in which the mapping of Dirac gaugino masses becomes important. In particular in the so-called free-magnetic phase, an asymptotically free electric $SU(\nc)$ theory with $\nf$ flavours of 
quarks and a chiral adjoint $X$ with superpotential $W\supset \kappa X^{k+1}$,  
flows to an IR-free $SU(\nnc)$ theory with $\nf$ flavours of magnetic quark and a chiral adjoint, $x$ with superpotential $W\supset \kappa x^{k+1}$. The question is how would a Dirac mass in the electric theory manifest itself the IR magnetic theory?

Various techniques have been developed to map soft-terms in ${\cal N}=1$ SUSY \cite{Karch1998,Evans1995,Aharony1995a,Evans1995a,Giudice1998a,Arkani-Hamed1998,Cheng1998,Arkani-Hamed1998a,Kobayashi2000,Nelson2001,Luty1999,Abel2011}. It is well known for example, that one can recover the RG flow of a Majorana gaugino mass by expressing it as a spurion contribution to the holomorphic gauge coupling \cite{Giudice1998a,Arkani-Hamed1998,Arkani-Hamed1998a,Luty1999}
\begin{equation}
{\cal L} \supset \int d^2\theta \,S\,{\WW}^2 + \hc,\qquad S=\frac{1}{2g_S^2 } -\frac{i \Theta}{16\pi^2}  + \theta^2 \frac{m_{\lambda S}}{g_S^2},
\end{equation}
where the physical gauge coupling and masses are functions of $S+S^\dagger $ (and real normalisation superfields ${\cal Z}$). The fact that one can construct a holomorphic RG invariant 
\begin{equation}
\Lambda_S=\mu \exp \left( -\frac{16\pi^2 S(\mu)}{b} \right) 
\end{equation} 
where $b=3t_G-\sum _r t_r$ is the usual beta function coefficient, shows that the quantity 
\begin{equation}
\frac{m_{\lambda S}}{g_S^2} = -\frac{b}{16\pi^2 } \left[ \ln \Lambda_S \right]_\theta^2 
\end{equation} 
is preserved. In other words the gaugino mass can be understood perturbatively as arising from $F$-terms in the threshold contributions to the one-loop beta function, but because it is related to a holomorphic invariant of the RG-flow, 
one can argue that this ratio is also mapped through any non-perturbative regions of strong coupling. In particular, when an asymptotically free theory flows to an IR-free magnetic description we can deduce, 
$$\lim_{\mu\rightarrow\infty}  
\frac{m_{\lambda S}}{g_S^2} = 
\lim_{\mu\rightarrow 0}  \frac{\tilde m_{\tilde{\lambda} S}}{\tilde{g}_S^2}$$ where tilde's represent the quantities in the dual description. Similar treatments are possible for the squark masses by constructing invariants involving the field renormalisation superfields ${\cal Z}(\mu)  $. An alternative method is that in \cite{Abel2011} where the Majorana mass is related through the ABJ anomaly to the anomalous trace current, which is in turn related to the $R$-current. If the latter is broken only by the gaugino masses themselves, one obtains a mapping up to corrections suppressed by factors of $m_\lambda^2/\Lambda^2$. 

Unfortunately similar techniques are not instantly available for Dirac masses. The operator that would generate the Dirac mass is 
\begin{equation}
W \supset \frac{X \WW^\alpha \WW'_\alpha}{M} , 
\label{diracterm2} 
\end{equation}
where the effective Dirac mass is $m_D= D/M$. Unlike the gauge coupling, the non-spurion part of this non-renormalizable term is not one that was in the theory before we required it for the Dirac mass. Likewise there is no equivalent to the conserved $R$-current technique of \cite{Abel2011}.

However, because the SUSY breaking is supersoft, one can establish the lack of anything other than field and one-loop gauge coupling renormalisation 
to all orders in perturbation theory, which implies that 
\begin{equation}
\frac{\beta_{m_D}}{m_D} = \frac{\gamma_X}{2}+\frac{\beta_g}{g} \, ,
\label{pertmap}
\end{equation}
where $\gamma_X$ is the anomalous dimension of the adjoint (ESP) field $X$ \cite{Weinberg1998,Jack1999,Fox2002,Goodsell2012}\footnote{We define $\gamma_X=-\partial \ln Z_X/\partial t$ so that 
dim$(X) = 1+\gamma_X/2$ -- hence the factor of 1/2 compared to these references.}. If the theory contains a superpotential term $W\supset \kappa x^{k+1}$, then we can always trade $\gamma_X$ for $\beta_\kappa$. By definition (and non-renormalization) we have  $\kappa^{-1}\beta_{\kappa}= \frac{k+1}{2} \gamma_X$. eq. \ref{pertmap} can then be solved to give an RG invariant,  
$m_D/g\kappa^{\frac{1}{k+1}}$.
Therefore it seems reasonable to suppose that Dirac masses in an asymptotically-free UV SQCD+$X$ theory 
are mapped directly to Dirac masses in an IR-free SQCD+$x$ theory as 
\begin{equation}
\lim_{\mu\rightarrow\infty}  
\frac{m_{D}}{g \kappa^{\frac{1}{k+1}}} = 
\lim_{\mu\rightarrow 0}  \frac{\tilde{m}_{{D}}}{\tilde{g} \tilde{\kappa}^{\frac{1}{k+1}}}.
\label{proposal}
\end{equation}

The purpose of the present paper is to establish this map. As we mentioned above, in SQCD+$X$ there is no RG-invariant that can be built from the couplings of the ${\cal N}=1$ theory which yields the Dirac mass as a spurion. Therefore the mapping cannot be done directly. However within ${\cal N}=2 $ theories it {\em is} possible to map Dirac masses, as discussed in \cite{Luty1999}. There $X$ becomes part of the ${\cal N}=2$ gauge supermultiplet, ${\cal A}$, with the Yang Mills lagrangian arising from the canonical prepotential 
${\cal L}\supset  \int d^2 \theta_1 d^2 \theta_2 \Sigma {\cal A}^2 $ where the indices label the two thetas of ${\cal N}=2$ in some basis. Both the Dirac and Majorana gaugino masses 
can be generated from spurions in the chiral ${\cal N}=2$ superfield $\Sigma$, out of which an RG-invariant {\em can} be constructed. 

Our task therefore is to extend this mapping to the ${\cal N}=1$ SQCD+$X$ theory. This is a much more difficult proposition than it might at first seem, because of the so-called \emph{2 into 1 won't go} theorem of \cite{Cecotti1986,Cecotti1984}. Ideally one would like to first deform the ${\cal N}=2$ theory to ${\cal N}=1$ SQCD+$X$, and then add a second deformation for the Dirac (and Majorana) masses. But the theorem of \cite{Cecotti1986,Cecotti1984} greatly restricts the form that breaking of ${\cal N}=2$  to ${\cal N}=1$ can take: essentially it has to be driven through a combination of electric and magnetic Fayet-Iliopoulos (FI) terms as shown by Antoniadis, Taylor and Partouche (ATP) \cite{Antoniadis1996,Antoniadis1996a,Partouche1997} 
in a mechanism inspired by \cite{Seiberg1994b,Seiberg1994c}.   In particular the gauge coupling between the quarks and the adjoint field, which in the ${\cal N}=1$ language is  ${\cal L}\supset  \tilde{Q} X Q$ (and which is not present in SQCD+$X$), cannot be removed by FI-terms. It can at best be made inconsequential by generating a holomorphic mass for $X$ and reducing the theory to simple ${\cal N}=1$ SQCD -- without the $X$. 

We are therefore forced to proceed by the following circuitous route. We consider the $N_f=2N_c$ version of the ${\cal N}=1$ $SU(N_c)$ SQCD+$X$ theory. As well as the necessary $W\supset \kappa X^{k+1}$ operator, the theory is deformed with the operator $h\tilde{Q} X Q$ -- where $h\ll g$ is parametrically small. We show perturbatively that for $k=2$ this theory can flow to the ${\cal N}=2$ fixed line in the IR, where $h\rightarrow g $  and $\kappa\rightarrow 0 $. Therefore we arrive at an ${\cal N}=2$ theory deformed by an operator $W\supset \kappa X^{k+1}$ where now $\kappa$ is parametrically small. We then show that the $h$ coupling in the magnetic description of the deformed SQCD+$X$ induces the correct higgsing for any $k$, causing the dual theory to flow to the dual ${\cal N}=2$ theory. This provides some evidence that the behaviour persists at strong coupling.

Once we have seen how to flow to ${\cal N}=2$ duality with parametrically small ${\cal N}=1$ deformations, we next establish that those deformations can be generated by electric and magnetic FI-terms in an ${\cal N}=2$ theory with an appropriate prepotential (unfortunately necessitating the paraphernalia of harmonic superspace). We thus complete a route that allows us to go from an electric ${\cal N}=1$ $SU(N_c)$ SQCD+$X$ theory to its magnetic dual via an intermediate pair of ${\cal N}=2$ duals. Dirac masses can now be added into the theory by further FI-deformations but now they {\em can} be mapped directly across the ${\cal N}=2$ duality, and  then tracked down the dual RG-trajectories to the dual SQCD+$X$ theories using eq. \ref{pertmap}. A schematic of the overall picture (before adding soft terms) is shown in figure \ref{fig:flow}.
The conclusion is that the proposed mapping in eq. \ref{proposal} seems to be correct. 
\begin{figure}
\begin{center}
\includegraphics[scale=0.75]{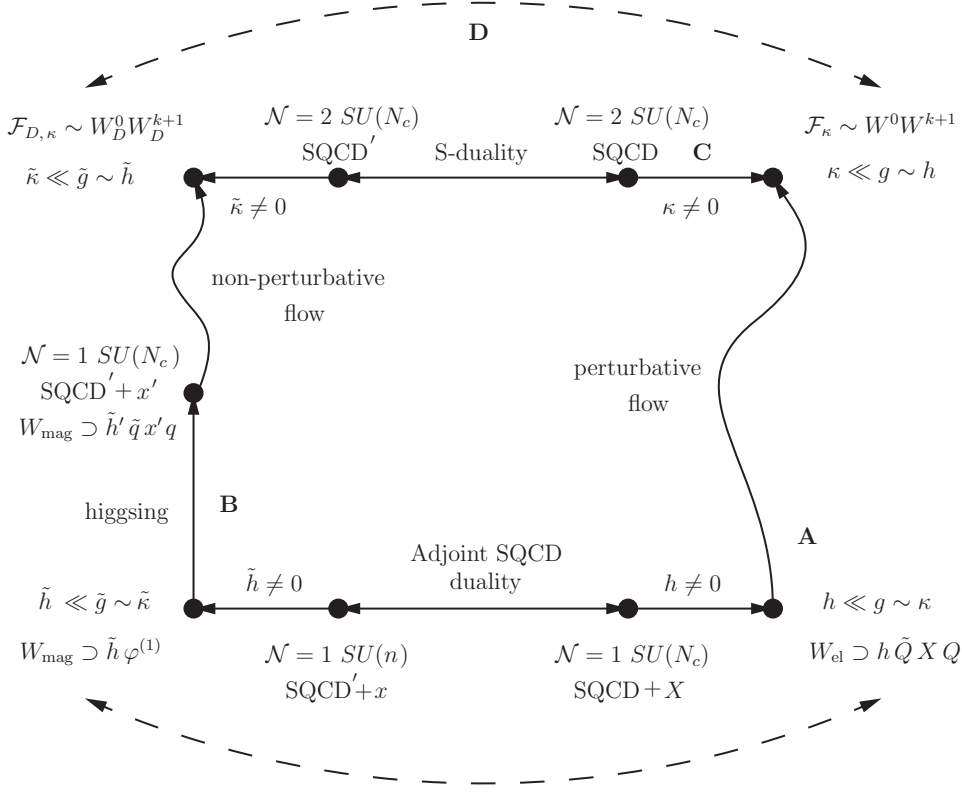}
\end{center}
\caption{\em The flow between $\NN=2$ S-duality and $\NN=1$ SQCD+$X$ duality. \textbf{A}: The duality of \cite{Kutasov1995} is deformed with a parametrically small $\NN=2$ quark gauge interaction. The resulting perturbative flow to $\NN=2$ SQCD is analysed in section \ref{sec:pertflow}. \textbf{B}: The the magnetic dual of the $\NN=2$ quark gauge interaction is observed to higgs the magnetic theory down to a gauge group of the same rank as the electric theory. This theory then flows to $\NN=2$ SQCD\tprime, as discussed in section \ref{sec:higgsing}. \textbf{C}: The electric theory of \cite{Kutasov1995} is now written as an $\NN=2$ theory broken to $\NN=1$ at low energies by electric and magnetic FI terms, as discussed in section \ref{sec:breakinginHSS}. \textbf{D}: By considering the Eguchi-Yang equations \cite{Eguchi1995}, the existence of a small dual $x^{k+1}$ deformation is shown to be required in the presence of a small electric $X^{k+1}$ deformation.}
\label{fig:flow}
\end{figure}

\section{From $\mathcal{N}=1$ SQCD+$X$ to $\mathcal{N}=2$ duality}

The programme outlined above naturally splits into two parts. The first -- the subject of this Section -- 
is to understand the RG flow from dual $\mathcal{N}=1$ SQCD+$X$ theories to dual $\mathcal{N}=2$ theories.
The second part is to investigate the induction in the latter of $\mathcal{N}=1$ deformations
and Dirac masses through FI terms, and also to determine explicitly how they map. As mentioned this 
requires some harmonic superspace technology, so it is postponed to the following Section and Section \ref{sec:dualrel}. 

Ideally one would like to be able to study the flow from SQCD+$X$ at a fixed point 
to $\NN=2$ SQCD at a fixed line. If this were possible one would be able to make general statements 
about the flow over the whole of the conformal window because one would know the anomalous dimensions precisely. 
In particular one might imagine that one could have a Banks-Zaks like fixed point for the Kutasov theory with 
a parametrically small deformation $W_{\mathrm{el}}\supset h\,\tilde{Q}XQ$ and with $\nf =2\nc$. Such a theory could conceivably flow to the ${\cal N}=2 $ theory at the fixed line. 
Unfortunately we shall begin by showing quite generally that flow between fixed points (lines) in these two theories is not possible. Either the ${\cal N}=1$ theory or the ${\cal N}=2$ theory cannot be at a fixed point (line). 

However in Subsection \ref{sec:pertflow} we examine the perturbative $k=2$ case (with $\nf=2\nc$), and find that there one can flow from one theory {\em not} at a fixed point (line) to the other at a fixed line (point). We then identify the higgsing mechanism whereby the (strongly coupled) dual ${\cal N}=1$ theory flows to the dual ${\cal N}=2$ theory; the fact that the higgsing in the dual descriptions would occur correctly for any $k$, suggests that the same flow would happen even in strongly coupled regimes.
At the end of this section we propose a means whereby (via decoupling) one might try more rigorously to extend the 
study to regions of parameter space in which neither dual is perturbative. 

\subsection{Non-perturbative generalities: regions of validity}

First let us make some general observations about the flow between $\mathcal{N}=1$ SQCD+$X$ and $\mathcal{N}=2$ theories. In particular one would like to know if it is possible to reach the fixed point of one theory by flowing from the fixed point of the other. As we shall see this is not possible. 

The general theory of interest is the ${\cal N}=1$ SQCD+$X$ duality of ref.~\cite{Kutasov1995} deformed by an
additonal $\tilde{Q} X Q $ coupling: 
\begin{equation}
W_{\mathrm{el}}=h\,\tilde{Q}XQ+\frac{\kappa}{k+1}\,\textrm{tr}_G\,{X^{k+1}},\label{eq:deform}
\end{equation}
where $X$ is the chiral adjoint field of the $\SU{\nc}$ gauge group.
The content and global symmetries of the $\mathcal{N}=1$ model with
no superpotential are given in table \ref{tab:KSSem}.
\begin{table}[ht]
\begin{center}
\begin{tabular}{c|c|c|c|c|c}
 & $SU(N_c)$ & $SU(N_f)_L$ & $SU(N_f)_R$ & $U(1)_B$ & $U(1)_R$ \\ \hline
$Q$ & \tfun & \tfun & \ttriv & $\frac1{N_c}$ & $1-R_X\frac{N_c}{N_f}$ \\
$\Qt$ & \tafun & \ttriv & \tafun & $-\frac1{N_c}$ & $1-R_X\frac{N_c}{N_f}$\\
$X$ & \tAd & \ttriv & \ttriv & 0 & $R_X$
\end{tabular}
\end{center}
 \caption{\em The matter content of the electric SQCD+$X$ model. All
the flavour charges are anomaly-free with respect to the gauge symmetry.
In the $W\sim X^{k+1}$ SQCD+$X$ $R_X=\frac{2}{k+1}$.}
\label{tab:KSSem}
\end{table}

The $h=g$ and $\kappa=0$ model corresponds to $\mathcal{N}=2$ SQCD,
while $h=0$ corresponds to the pure SQCD+$X$ model of \cite{Kutasov1995}.
The mapping of the beta function under duality, suggests that SQCD+$X$ has a conformal window for 
\begin{equation}
\frac{1}{k-\frac{1}{2}}\nc<\nf<2\nc,
\end{equation}
and is in the free magnetic phase for 
\begin{equation}
\frac{1}{k}(\nc+1)<\nf\leq\frac{1}{k-\frac{1}{2}}\nc.
\end{equation}
However the phase space is complicated by the fact that mesons (the first being $M=\tilde{Q}Q$) can sequentially 
decouple as their anomalous dimension hits unity. This was studied in ref.~\cite{Kutasov2003}. To summarise the main results from that work, in order to avoid this in the theory without the superpotential, 
one requires $\nf > \nc /(3+\sqrt{7})$. More restrictive is that, in the theory with $h=0$, the operator $X^{k+1}$ is only relevant 
when $\nf < \nc /x_k$ where $x_k = \sqrt{\frac{1}{20}  \left( \frac{(5k-4)^2}{9} +1 \right) }$. Thus for example for $k=2$ this corresponds to 
$\nf < 2 \nc $, with higher $k$ requiring successively smaller $\nf$ for the operator to be relevant. 

In the present context we are envisaging flowing from this theory to 
 the $\mathcal{N}=2$ theory with small $\kappa$ induced by an FI
term, and so we are interested in the influence of the operator $h$, and 
anticipate that the RG flow will be dominated by either $h$ or $\kappa$
in different regions. Therefore these bounds cannot be immediately used to draw
conclusions about the flow in the present context, but they serve as a useful guide in the different regions. 

Instead we can dismiss (by contradiction) the possibility of flow from fixed-point to fixed-point (line) using the $a$-theorem, as follows. 
Defining the dimensionless coupling $\eta_\kappa = \kappa \, \mu^{k-2}$, 
the supersymmetric RG equations are given to all orders by 
\begin{align}
\frac{dg^{2}}{dt} & = 2\,g\beta_{g}, \qquad &
\frac{dh^{2}}{dt} & = h^{2}(\gamma_{X}+2\gamma_{Q}), \qquad &  
\frac{d\eta_{\kappa}^{2}}{dt} & =  \eta_{\kappa}^{2}\left[(k+1)(\gamma_{X}+2)-6\right],\nonumber \\
\noalign{\[
\beta_{g} = -\frac{g^{3}}{16\pi^{2}}\,\frac{3\,C_{2}(\Ad)-2\,\nf \,T(\fun)(1-\gamma_{Q})-T(\Ad)(1-\gamma_{X})}{1-T(\Ad)\,\tfrac{g^2}{8\pi^2}},\]}
T(\fun) & = \frac{1}{2}, & C_{2}(\fun) & = \frac{\nc^{2}-1}{2\nc}, & C_{2}(\Ad) & = \, T(\Ad) = \nc,
\end{align}
where the first line is by definition, and where $\beta_g$ is the all orders NSVZ beta function. 
%At the BZ fixed point where $h$ is marginal, $2R_{Q}+R_{X}=2$ so
%that $R(\tilde{Q}XQ)=2$ as required. More generally
Assume that both theories {\em can} be at a fixed point with the same values of $\nc$ and $\nf$. Then the vanishing
of the NSVZ $\beta$-function (using $\gamma=3R-2$ at a fixed-point) agrees with the $R_{Q}$-charges shown
in table \ref{tab:KSSem} which are determined from absence of mixed
$SU(\nc)^{2}\times\U1_{R}$ anomalies. Unless $\nf=2\nc$ precisely,
the values of $R_{X}$ consistent with the $h$ or $\kappa$ coupling
are $R_{X}=0$ and $R_{X}=2/(k+1)$ respectively, so that $h$
breaks the $R$ symmetry of SQCD+$X$: therefore if $\nf\neq2\nc$ there
can be no fixed point behaviour unless either $h$ or $k$ are zero.
If and only if $N_{f}=2N_{c}$, can one find fixed-point solutions
of the RGEs with non-zero $h$ and $\eta_{\kappa}$. They are
at $\gamma_{Q}=(-2+k)/(1+k)$ and $\gamma_{X}=(4-2k)/(1+k)$ which
 correspond to the required values for the superconformal
$R$-symmetry at $N_{f}=2N_{c}$, namely 
\begin{equation}
R_{Q} = 1-\frac{1}{k+1},\qquad
R_{X} = \frac{2}{k+1}.
\end{equation}
But because $h,\kappa$ preserve precisely the same $R$-symmetry
with $R$-charges completely constrained, the $a$-theorem \cite{Komargodski2011,Intriligator2004,Intriligator2003} now tells
us that SQCD+$X$ at a fixed-point cannot flow to the $\mathcal{N}=2$
fixed line (otherwise the flow would occur without any decrease in $a$). We
conclude that both theories cannot be at a fixed point with the same values of $\nc$ and $\nf$. 

\subsection{Perturbative flow to $\NN=2$ SQCD}
\label{sec:pertflow}
We are therefore forced to consider the next best option, which is for the theory to flow from a {\em third} fixed point to the ${\cal N}=2$  fixed line, via the ${\cal N}=1$ Kutasov theory. As the intermediate Kutasov  theory is not at a fixed point we can obviously say less in this regime: we first resort to a perturbative analysis in which $k=2$ and $\nf=2\nc$, since that theory is weakly coupled (note: since all the anomalous dimenions are small, we do not have to worry about particles decoupling in this regime). The third fixed point we start from is SQCD+$X$ with no superpotential, so one might  imagine that $\kappa$ would just be marginally irrelevant. However the theory exhibits quasi-fixed point behaviour: $\kappa/g$ runs to a fixed value even though $g$ itself is running at two loops. (Note that conversely by choosing $\nf \lesssim 2\nc$ one can flow from the Banks-Zaks fixed point of the ${\cal N}=1$ theory to an ${\cal N}=2$ theory with quasi-fixed $h$ coupling.) 

We see this explicitly by using the perturbative anomalous dimensions; 
\begin{align}
\gamma_{Q}=\gamma_{\tilde{Q}} & = \frac{1}{4\pi^{2}}\,C_{2}(\fun)\left(h^{2}-g^{2}\right),\nonumber \\
\gamma_{X} & = \frac{1}{4\pi^{2}}\left[\nf T(\fun)\, h^{2}+\delta_{k,2}\left(4\,C_{2}(\fun)-\frac{3}{2}\,T(\Ad)\right)\,\eta_{\kappa}^{2}-C_{2}(\Ad)\, g^{2}\right],
\end{align}
Note that there is a $\eta_{\kappa}^{2}$
contribution to $\gamma_{X}$ at one-loop order for $k=2$, but at
two loop order for $k=3$ and successively higher order for higher
values of $k$. Starting in the UV with a weakly coupled theory with 
$\eta_{\kappa}$ and $h$ arbitrarily small but $h\ll\eta_{\kappa}$, one finds that the theory first flows
to $\gamma_{X}=0$ and ${\eta_{\kappa}^{2}}/{g^{2}}=\frac{2C_{2}(\Ad)}{8C_{2}(\fun)-3T(\Ad)}$, 
with $g$ experiencing two-loop running. Eventually, $h$ turns on
and the theory flows to the $\mathcal{N}=2$ theory with $\eta_{\kappa}$ flowing
to zero. A numerically solved example is shown in figure \ref{fig:flow}.

\begin{figure}[h]
\includegraphics[scale=0.5]{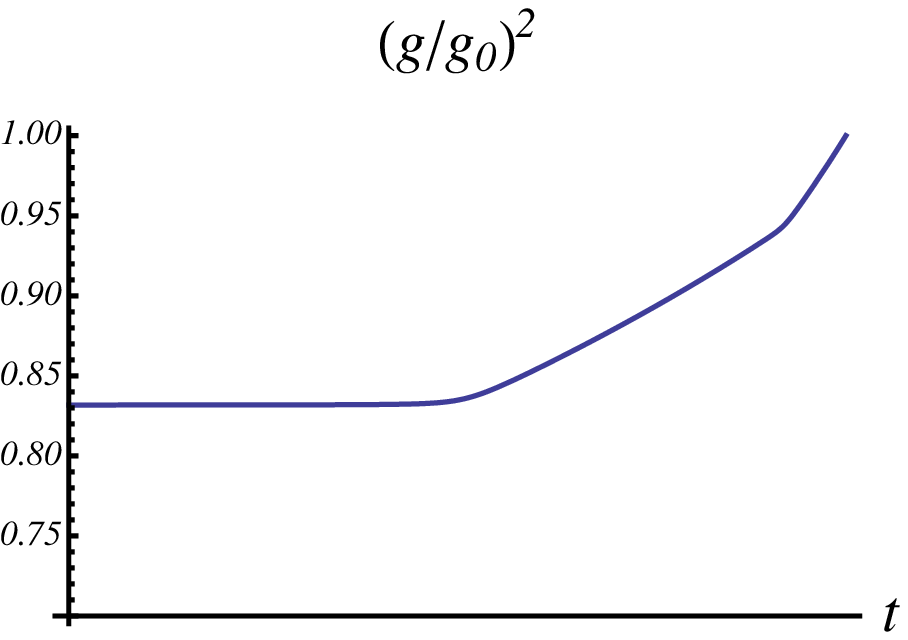}~\,~\includegraphics[scale=0.5]{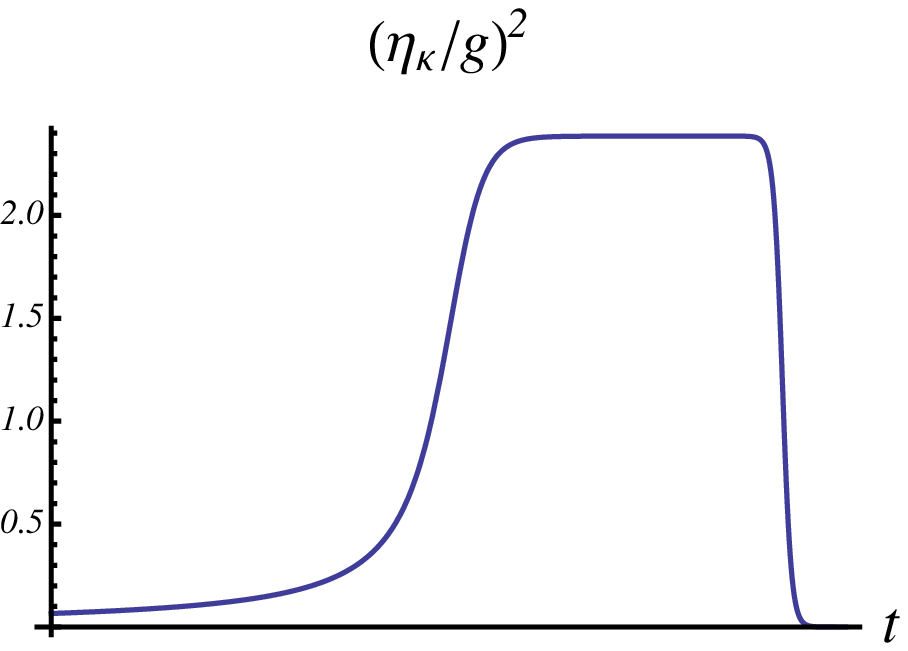}~\,~\includegraphics[scale=0.5]{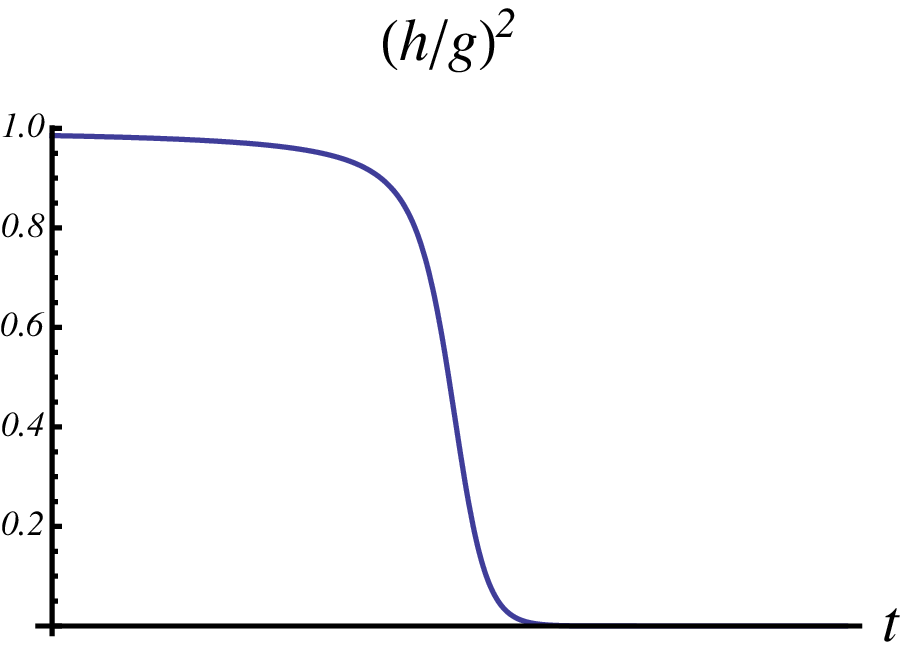}
\caption{RG flow of $g$, $\eta_\kappa$ and $h$ from the UV (right) to the (IR) left. The horizontal axis is $t=\log\,\mu$, and we take $N_c=5$, $N_f=10$.}
\label{fig:flow}
\end{figure}

\subsection{Higgsing in the dual theory and flow to $\NN=2$ SQCD\tprime}
\label{sec:higgsing}
Having demonstrated that RG-flow between the two theories is at least possible if one of them is at a fixed point, we now ask how such a 
flow appears in the dual description. 
In particular while the dual of the $\mathcal{N}=2$ theory is also
an $SU(N_{c})$ gauge theory, the dual of SQCD+$X$ is an $\SU{\nnc}=\SU{k\nf-\nc}$
gauge theory: i.e. for $k=2$ it is an $SU(3\nc)$ theory. We will now show that 
it is the growing $h$ coupling which induces the necessary breaking
$\SU\nnc\rightarrow \SU\nc$ in the IR of the magnetic dual description. Moreover 
the breaking is correct for any value of $k$.  We take this to be evidence that the same flow would occur even when both the theory and its dual are strongly coupled although we cannot show that they always end up at the $\NN=2$ fixed line.

Let us see this in detail. The spectrum of the magnetic SQCD+$X$ theory
has mesons denoted $m_{j}$ identified as; 
\begin{equation}
m^{(j)}=\tilde{Q}X^{j-1}Q,\hspace{5mm}j=1\ldots k,\label{eq:meson-1}
\end{equation}
with canonically normalized fields $\varphi^{(j)}\sim\Lambda^{-j}m^{(j)}$.
The field content of the magnetic theory is $q$, $\tilde{q}$, $\vphi^{(j)}$
and $x$, where $x$ is an adjoint of the $\SU{\nnc}=\SU{k\nf-\nc}$
magnetic gauge group, as summarised in Table \ref{tab:KSSmm-1}. 

\begin{table}[ht]
\begin{center}
\begin{tabular}{c|c|c|c|c|c}
 & $SU(n)$ & $SU(N_f)_L$ & $SU(N_f)_R$ & $U(1)_B$ & $U(1)_R$ \\ \hline
$q$ & \tfun & \tafun & \ttriv & $\frac1n$ & $1-R_x\frac{n}{N_f}$ \\
$\qt$ & \tafun & \ttriv & \tfun & $-\frac1n$ & $1-R_x\frac{n}{N_f}$\\
$x$ & \tAd & \ttriv & \ttriv & 0 & $R_x$\\
$\vphi^{(j)}$ & \ttriv & \tfun & \tafun & 0 & $2-2\,R_x\,\frac{N_c}{N_f}-R_x(j-1)$
\end{tabular}
\end{center}
\caption{\em The matter content of the magnetic theory in the dual SQCD+$X$
model; $n=kN_{f}-N_{c}$.\label{tab:KSSmm-1}}
\end{table}

In general the magnetic superpotential is 
\begin{equation}
W_\textrm{mag}=h\,\vphi^{(2)}_{nm}\de_{nm}+\frac{\tilde\kappa}{k+1}\,\tr_G(x^{k+1})+\sum_{j=1}^k\tilde{c}_j\,\vphi^{(j)}_{nm}\,\tr_G(\qt_mx^{k-j}q_n)
\label{eq:wmag-1}\end{equation}
where $n,m$ are flavour indices,
and for the special case $k=2$ we have (dropping the indices)  
\begin{equation}
W_{\mathrm{mag}}=h\varphi^{(2)}+\frac{\tilde\kappa}{3} x^{3}+\left(\tilde{c}_{1}\varphi^{(1)}\tilde{q}xq+\tilde{c}_{2}\varphi^{(2)}\tilde{q}q\right).\label{eq:wmag-1-1}
\end{equation}
Not surprisingly the SU(3$\nc$) dual theory is at the strongly coupled
boundary of the conformal window, 
\begin{equation}
\frac{1}{k-\frac{1}{2}}\,\nnc\,=2\nc=\nf.\label{eq:nfrange-1-1}
\end{equation}
Because the anomalies must match, the $a$-parameters of the electric
and magnetic theories also match at the endpoints of the flow, so
thanks to the $a$-theorem, even though we cannot numerically solve the RG equations
in the dual description, we know that it too flows to the $\mathcal{N}=2$
theory in the IR. Therefore in the UV we expect a strongly coupled theory
with $\tilde{h}\propto \tilde{c}_1=\tilde{\kappa}=0$, which we expect to flow to an intermediate strongly
coupled $SU(3\nc)$ SQCD+$X$ theory with adjoint coupling $\tilde{h}=0$, and
thence to an $SU(\nc$) theory with adjoint coupling $\tilde{h}=\tilde{g}$. 

Indeed the $\varphi^{(2)}$ eq. of motion sets 
\begin{equation}
\tilde{c}^{(2)}\tilde{q}q=-h.
\end{equation}
These equations have rank $\nf=2\nc$ and thus, once it turns on,
the coupling $h$ induces the required higgsing SU$(3\nc)\hookrightarrow$\,SU$(\nc)$.
By using colour and flavour rotations, we can arrange the VEVs for
the magnetic quarks in a form that makes explicit the $\nc\times\nc$
blocks: 
\begin{equation}
q=\tilde{q}=-\sqrt{\frac{h\Lambda^{2}}{\tilde{c}^{(2)}}}\left(\begin{array}{cc}
\mathbb{I}_{\nc\times\nc} & \cdot\\
\cdot & \mathbb{I}_{\nc\times\nc}\\
\cdot & \cdot
\end{array}\right).
\end{equation}
Writing the SU($3\nc$) adjoints as 
\begin{equation}
x=\left(\begin{array}{cc}
z & y\\
\tilde{y} & \hat{x}
\end{array}\right)
\end{equation}
where $z$ is $2\nc\times2\nc$ and $\hat{x}$ is $\nc\times\nc$,
the $\tilde{c}_{1}$ coupling then becomes an effective mass term for the
adjoint $z$ and the traceless mesons $\bar{\varphi}^{(1)}=\varphi^{(1)}-\frac{1}{2\nc}\mbox{tr}(\varphi^{(1)})$,
of the form
\begin{equation}
-\frac{h\tilde{c}_{1}}{\tilde{c}_{2}}\bar{\varphi}^{(1)}z.
\end{equation}
Note that colour-flavour is broken to the diagonal, SU$(3\nc)\times\SU\nf\hookrightarrow\mbox{SU}(\nc)\times\mbox{SU}(\nf)_{D}$,
and this term represents a Dirac mass for two adjoints of the remaining
diagonal flavour group. In addition $\varphi^{(2)}$ gets a mass together
with the higgsing $2N_{c}$ block of $q$. In detail writing $q=\left(\begin{array}{c}
v+\eta\\
\rho
\end{array}\right)$ and $\tilde{q}=\left(\begin{array}{c}
v+\tilde{\eta}\\
\tilde{\rho}
\end{array}\right)$, we find a mass term term $W\supset \tilde{c}_{2}\left(\eta+\tilde{\eta}\right)\varphi^{(2)}v$,
with the $8\nc^{2}$ massless $\eta-\tilde{\eta}$ Goldstone modes
being eaten by the $8\nc^{2}$ heavy gauge bosons of the broken SU$(3\nc)$.
Meanwhile $\rho,\tilde{\rho}$ are the light quarks of the remaining
unbroken $\SU\nc$. The superpotential for the remaining effective
$\SU\nc$ theory is 
\begin{equation}
W_{\mathrm{mag}}=\frac{\tilde{\kappa}}{3} x^{3}+\tilde{h}\tilde{\rho}x\rho
\end{equation}
where $\tilde{h}=\tilde{c}_{1}\mbox{tr}(\varphi^{(1)})$ is dynamical in the dual theory.
As stated above, the $a$-theorem tells us that this SQCD+$X$ theory flows to the 
dual $\mathcal{N}=2$ fixed point.

It is straightforward to extend the above discussion to arbitrary $k$, to check that the $h$ coupling induces the required breaking 
$SU((2k-1)\nc)\times\SU\nf\hookrightarrow {SU}(\nc)\times {SU}(\nf)$. 
From eq. \ref{eq:wmag-1-1} we find that the $X$ and $\vphi$ equations of motion are
\begin{align}
\vphi^{(j)}:&\quad0=h\,\de_{nm}\,\de_{2j}+\tilde{c}_j\,\tr(\qt_mx^{k-j}q_n)\\
x:&\quad0=\tilde\kappa\,x^k+\sum_{j=1}^k\tilde{c}_j\,\vphi^{(j)}_{nm}\sum_{r=0}^{k-j-1}x^{k-j-1-r}\,q_n\,\qt_m^T\,(x^r)^T\, .
\end{align}
From the first condition we see for $k\geq3$ and non-zero $\tilde{c}_j$
\begin{align}
\tr\vev{\qt_mx^{k-1}q_n}=\tr\vev{\qt_mx^{k-3}q_n}=\ldots=\tr_G\vev{\qt_mxq_n}=\tr\vev{\qt_mq_n}&=0\label{eq:cond1}\\
\tr\vev{\qt_mx^{k-2}q_n}&\neq0.\label{eq:cond2}
\end{align}
Let us write $x, q$ and $\qt$ as
\begin{equation}
x=\begin{pmatrix}z&y\\ \tilde{y}&\hat x\end{pmatrix},\quad q=\begin{pmatrix}v+\eta\\ \rho_1\\ \rho_2\end{pmatrix},\qquad \qt^T=\begin{pmatrix}\tilde{\rho}_1\\v+\tilde\eta\\\tilde{\rho_2}\end{pmatrix},
\end{equation}
where $z$ is an $(k-1)N_f\times(k-1)N_f$ matrix, $v,\eta,$ and $\tilde\eta$ are $N_f\times N_f$ matrices, $\rho_1$ and $\tilde{\rho}_1$ are $(k-2)N_f\times N_f$ matrices, and $\rho_2$ and $\tilde{\rho}_2$ are ${N_c}\times N_f$ matrices.
We can solve equations \ref{eq:cond1} and \ref{eq:cond2} by taking $z$ as
\begin{equation}
\vev{z}\sim
\begin{pmatrix}
0_{N_f\times N_f} & \mathbb{I}_{N_f\times N_f}&\cdot&\cdot\\
\cdot& \cdot & \ddots&\cdot\\
\cdot& \cdot & \cdot & \mathbb{I}_{N_f\times N_f}\\
\cdot & \cdot & \cdot & 0_{N_f\times N_f}
\end{pmatrix}\end{equation}
such that
\begin{equation}
\vev{z^{k-2}}\sim
\begin{pmatrix}
\;\;\cdot\;\; & \mathbb{I}_{N_f\times N_f}\\
\cdot& \cdot
\end{pmatrix}\end{equation}
and then seperating the VEVs of $q$ and $\qt$ by $k-2$ permutations,
\begin{equation}
\vev{\qt}\sim\begin{pmatrix}\mathbb{I}_{N_f\times N_f}\\ \cdot\end{pmatrix},\qquad \vev{q}\sim\begin{pmatrix}0_{(k-2)N_f\times N_f}\\ \mathbb{I}_{N_f\times N_f}\\ \cdot\end{pmatrix}\, ,
\end{equation}
so that clearly
\begin{equation}
\vev{x^{k-2}q}\sim\begin{pmatrix}\mathbb{I}_{N_f\times N_f}\\ \cdot\end{pmatrix}\sim\vev{\qt},
\end{equation}
as required. Then $\langle z \rangle $ which is rank $(k-2)N_f$, together with $\langle q\rangle $, leave the bottom $\rho_2,\tilde{\rho}_2$ block and hence $SU(\nc)\times {SU}(\nf)$ unbroken.

\subsection{Flow away from $N_f=2N_c$}
\label{sec:integratingout}
Having established this connection, one could devise ways to reach more general SQCD+$X$ configurations that have arbitrary $N_f$ and $N_c$. (We will leave a complete study for future work.) From our $\nf=2\nc$ electric theory we can add $\Delta$ additional heavy quarks $Q',\tilde{Q}'$ with mass terms $W\supset m_{Q'} Q'\tilde{Q}'$ with $m_{Q'}$ being chosen to be in the SQCD+$X$ period of running. Instead of running to a free field theory, the original electric theory now heads towards a Landau pole in the UV. Meanwhile in the magnetic dual description, the mass term becomes a linear term for the new meson $\varphi' = \Lambda^{-1} Q'.\tilde{Q}'$, which induces a higgsing for the new magnetic quarks $\tilde{q}'.q' \sim m_Q'\Lambda $. This strongly coupled theory is asymptotically free in the UV. Conversely, as well as $Q'$ and $\tilde{Q}'$, one can also add a meson $\Phi$ into the electric theory together with a linear coupling $W\supset \mu^2 \Phi$: this implies a symmetry restoration SU$(\nc)\hookleftarrow \, $SU$(\nc+\Delta)$ at the scale $\mu$, and the theory can flow to the SQCD+$X$ conformal fixed point of that theory. (Note that the $R$-symmetry associated with this fixed point can be compatible with the previous ${\cal N}=2$ $R$-symmetry because we have integrated in more degrees of freedom.) Note that the mass deformations may be already introduced at the ${\cal N}=2$ level \cite{Sohnius1978,Ohta1986}.

\section{Breaking $\NN=2$ to $\NN=1$ with general $W(X)$}

\subsection{Overview}

We have shown that dual SQCD+$X$ theories deformed by $\NN=2$ operators can flow to dual $\NN=2$ theories deformed by $\NN=1$ operators. We will now study the $\NN=2$ duality itself. In particular, we will show that the residual $\NN=1$ deformations can be understood as being induced by the ATP mechanism, and that they map consistently into each other under the $\NN=2$ duality. The ATP mechanism was formulated in harmonic superspace (HSS) in \cite{Ivanov1997}, and has been coupled to a number of interesting theories \cite{Fujiwara2005,Fujiwara2005a,Fujiwara2004a,Fujiwara2004,Fujiwara2006,Fujiwara2006a,Itoyama2008}, of which the most relevant for this study is \cite{Fujiwara2005} where the the theory is $\NN=2$ SQCD. We will proceed as follows:
\begin{itemize}
\item In Subsection \ref{sec:N=2SQCD} we write $\NN=2$ SQCD in the HSS formalism. For comparison, we also write this theory in the standard $\NN=1$ superspace in Appendix \ref{sec:N=2SQCD,N=1SS}.
\item In Subsection \ref{sec:breaking2to1}, noting the restriction from the \emph{2 into 1 won't go} theorem \cite{Cecotti1986,Cecotti1984}, we collect the neccessary ingredients to acheive $\NN=2\rightarrow\NN=1$ breaking, and check that it successfully reproduces the usual ATP mechanism.
\item In \ref{sec:scalarpot} we show that a specific choice of the prepotential $\FF(\WW)$ generates the desired form of the scalar potential (in eq.~\ref{340}), matching the known result from $\NN=1$ superspace which is presented in Appendix \ref{sec:2to1N=1SS}.
\item In \ref{sec:gaugino-fermion} we show that the same prepotential generates the desired fermion lagrangian (in eq.~\ref{345}) also matching the known result from $\NN=1$ superspace.
\end{itemize}

\subsection{$\NN=2$ $SU(N_c)$ SQCD}
\label{sec:N=2SQCD}

The low energy effective action (LEEA) for $\NN=2$ $SU(\nc)$ SQCD is 
\cite{Gates1984}\footnote{HSS expansions for $Q^+, V^{++},$ and $\WW$ are
\begin{align}
Q^+(\zeta,u)&\supset Q^i(x_A)u_i^++\tta^+\psi_Q(x_A)+\ttab^+\psi_\Qt(x_A),\\
V^{++}(\zeta,u)&\supset i\sqrt2(\ttab^+)^2X(x_A)+4(\ttab^+)^2\tta^+\lda^i(x_A)\,u_i^--2i\tta^+\si^\mu\tta^+A_\mu(x_A)+6(\tta^+)^2(\ttab^+)^2D^A(x_A)(u^-u^-)^A,\\
\WW(\zeta,u)&\supset i\sqrt2\,X(x_A)-2\,\tta^+\lda^i(x_A)\,u_i^-+\tta^i\si^{\mu\nu}\tta_i\,F_{\mu\nu}(x_A)+2(\tta\tta)^AD^A(x_A).
\end{align}
The gauge covariant derivative $\DD^{++}$ and further HSS definitions are provided in Appendix \ref{sec:hhsnotation}. See \cite{Galperin2001,Galperin1985} for a review.
}
\begin{equation}
S^{\NN=2}_\textrm{QCD}=S^{\NN=2}_\textrm{SYM}+S^{\NN=2}_Q,
\label{eq:N=2SQCD}
\end{equation}
\begin{equation}
 S^{\NN=2}_\textrm{SYM}
=\frac1{16\pi i}\int d^4x\,(D)^4\FF(\WW)+\hc,
\qquad
S^{\NN=2}_Q=- \int\,du\,d\zeta^{-4}\,\Qt^+\,\DD^{++}Q^+,\label{eq:N=2SQCDparts}
\end{equation}
where $Q^+$ is a Fayet-Sohnius (FS) hypermultiplet \cite{Fayet1976,Sohnius1978}, $V^{++}$ is a $\NN=2$ vector multiplet, and $\WW$ is the full $\NN=2$ gauge field strength. Note that $\tilde{Q}^+$ in this equation refers to the antipodal~$\times$~hermitian conjugation of the $Q^+$ hypermultiplet and should not be confused with the $\NN=1$ superfield $\tilde{Q}$ (See Appendix \ref{sec:hhsnotation} for details). In addition we canonically normalise the hypermultiplets in contrast with the usual convention.  $\FF(\WW)$ is the prepotential, and is a gauge invariant function of only $\WW\equiv\WW^at_a$, $a=1,\ldots,N_c^2-1$, whose general form is
\begin{equation}
\quad\FF(\WW)
=
\sum_M\frac1{M!}
\sum_{m_1\,\ldots\, m_M}\,\frac{c_{m_1\ldots m_M}}{m_1!\ldots m_M!}\,\tr_G\,(\WW^{m_1})\ldots\tr_G\,(\WW^{m_M}),\label{eq:prepotential}
\end{equation}
where $\tr_G$ is a trace over the $SU(N_c)$ gauge indices, the $m_i$ represent powers and not gauge indices, and the coefficients $c_{m_1\ldots m_M}$ arise from integrating out microscopic degrees of freedom, and have been exactly determined in specific cases, for example in \cite{Seiberg1994b,Seiberg1994c}.  We define derivatives of the prepotential and the metrics\footnote{Note that by subscript-$a$ we will always mean $\partial /\partial \WW^a \equiv (i\sqrt{2})^{-1}\partial /\partial X^a $.}
\begin{equation}
\FF_{a_1\,\ldots\,a_N}(\WW)\equiv\frac{\partial^N\FF(\WW)}{\partial \WW^{a_1}\,\ldots\,\partial \WW^{a_N}},\qquad h_{ab}\equiv \textrm{Re}\,\FF_{ab}|,\qquad g_{ab}\equiv \textrm{Im}\,\FF_{ab}|,
\end{equation}
where $\OO|\equiv \OO(\tta=\ttab=0)$. 
\begin{table}[ht]
\begin{center}
\begin{tabular}{c|c|c}
 & $SU(N_c)$ & $SU(N_f)$\\ \hline
$Q^+$ & \tfun & \tfun
\end{tabular}
\end{center}
\caption{\em $\NN=2$ superfield representations in $\NN=2$ SQCD}
\label{tab:n=2fieldsa}
\end{table}
The resulting theory up to four derivatives in the prepotential is presented for completeness in Appendix \ref{sec:N=2HSS}.

%The most important terms for us will be the extra terms involving $D^{a,\,A}$,
%\begin{equation}
%\label{genferm}
%4\pi\LL_\textrm{fermion-D}=\frac i2\,\FF_{abc}|(\lda^a\lda^b)^AD^{c,\,A}|+\hc \, ,
%\end{equation}
%as well as four-fermion interactions and higher derivative terms that we do not write. Once FI-terms are introduced, eq. \ref{genferm} can be a source of chiral $\NN=1$ masses 
%and yukawa interactions, as well as $\NN=0$ gaugino masses. 
%
%In this section we will be mainly concerned with the classical extension to the theory that generates the correct $\NN=1$ deformations, and therefore we will often work by deforming around a canonical prepotential:  
%\begin{equation}
%\FF(W)=\frac {\tau_{ab}}{2} {W^aW^b}+\mathrm{deformation}. \label{eq:canpre2}
%\end{equation}
%This is sufficient to consider $\NN=1$ deformations of a weakly coupled $\NN=2$ theory where there is negligible RG-flow. The discussion of duality in section 4 will require more careful treatment of the leading part.

\subsection{$\NN=2\rightarrow\NN=1$ $SU(N_c)$ SQCD}

\label{sec:breaking2to1}

\subsubsection{Formulation in harmonic superspace: the ATP mechanism}

\label{sec:breakinginHSS}

To acheive spontaneous breaking of $\NN=2\rightarrow\NN=1$ via the ATP mechanism, we first extend the gauge theory $SU(N_c)\rightarrow SU(N_c)\times U(1)_\fn$, where $Q^+$ is charged under the $U(1)_\fn$ factor as shown in table \ref{tab:n=2fieldsb}.
\begin{table}[ht]
\begin{center}
\begin{tabular}{c|c|c|c}
 & $SU(N_c)$ & $U(1)_\fn$ & $SU(N_f)$\\ \hline
$Q^+$ & \tfun & 1 & \tfun
\end{tabular}
\end{center}
\caption{\em $\NN=2$ superfield representations in $\NN=2$ SQCD coupled to $U(1)_\fn$}
\label{tab:n=2fieldsb}
\end{table}
The resulting action is the same as in \ref{eq:N=2SQCD} and \ref{eq:N=2SQCDparts} with prepotential $\FF(\WW,\WW^\fn)$ written as a general expansion in $\WW$'s, 
and the covariant derivative
\begin{equation}
\DD^{++}=D^{++}+i\,(V^{++}+V_\fn^{++}).
\end{equation}
The ${}^{{}_{{}_{{}_\fn}}}$-index on $V^{++}_\fn$ or $\WW^\fn$ is equivalent to the trace $U(1)$ element of the $U(N_c)$ gauge group in \cite{Fujiwara2005}, in the sense that we can 
define a K\"ahler metric for the whole gauge theory through the prepotential $\FF_{a_1\,\ldots\,a_N}(\WW,\WW^\fn)$. From now on we use take following notation to distinguish $SU(N)$ and $U(1)$ indices
\begin{equation*}
\at=1,\ldots,N_c^2-1,\qquad a=\,\fn\,,1,\ldots,N_c^2-1,
\end{equation*}
and we normalize the $U(1)_\fn$ generator as $t_\fn=\frac1{\sqrt{2N_c}}\,\mathbb{I}_{N_c\times N_c}$. 

$\NN=2$ SUSY can be broken spontaneously by giving the electric and dual magnetic\footnote{see Section \ref{sec:dualrel}.} $D$ terms of the $U(1)_\fn$ gauge field a constant shift. The dual magnetic $D$ term $D_{D,\,\fn}^A$ is shifted by the electric FI term
\begin{equation}
4\pi\, S^{\NN=2}_{\textrm{El FI},\,\fn}
=\int du\,d\zeta^{(-4)}\,\xi^{++}(V^\fn)^{++}+\hc
=2\int d^4x\,\xi^AD^{\fn,\,A}\,+\hc \label{eq:elFI}
\end{equation}
where $\xi^{++}\equiv \xi^{ij}u_i^+u_j^+=2\,\xi^A(u^+u^+)^A$. This shift can be seen by writing the whole action as an integral over the analytic subspace and varying it with respect to $V_\fn^{++}$ yielding the equation of motion \cite{Zupnik1987,Ivanov1997}
\begin{equation}
(D^{+})^2\,\FF_\fn-\hc=4\,i\,\xi^{++}.\label{eq:D+eom}
\end{equation}
Because $\FF_\fn\equiv\WW_{D,\,\fn}\supset 2(\tta\tta)^A D_\fn^A$, the equation of motion \ref{eq:D+eom} shifts the magnetic dual $D$ term $D_{D,\,\fn}^A$ by an imaginary part on-shell \cite{Fujiwara2005}:
\begin{equation}
\DB^A_{D,\,\fn}=D^A_{D,\,\fn}+4\,i\,\xi^A,\qquad \DBb^A_{D,\,\fn}=\Db^A_{D,\,\fn}-4\,i\,\xib^A.
\end{equation}
Similarly, the electric $D$ term is shifted by an FI term for the dual magnetic gauge field, and turns out to be of the form
\begin{equation}
4\pi \,S^{\NN=2}_{\textrm{Mag FI},\,\fn}
=2\int d^4x\,\xi_D^A\left[(D)^4(\tta\tta)^A\left(\FF_\fn+\FF_{\fn\fn}\,4\,i\xi_D^B(\tta\tta)^B\right)-2\,\QQ_\fn^A\right]+\hc
\label{eq:magFI}
\end{equation}
where\footnote{our symmetrization conventions are $a^{(i_1\ldots i_n)}\equiv\frac1{n!}\left(a^{i_1\ldots i_n}+\textrm{permutations}\right)$.}
\begin{equation}
\label{implicitqhat}
\QQ_a^{ij}\equiv 4\pi \Qb^{(i}t_aQ^{j)}=-\bar{\QQ}_a^{ij}.
\end{equation}
For later reference the explicit form of the $\QQ^A$'s is 
\begin{equation}
\label{queues}
\frac{\QQ^1_a}{2\pi i}= -(\overline{Q^2}t_aQ^1 + \overline{Q^1}t_aQ^2)\, ; \;\; 
\frac{\QQ^2_a}{2 \pi}= \overline{Q^2}t_aQ^1  - \overline{Q^1}t_aQ^2\, ; \;\;
\frac{\QQ^3_a}{2\pi i}=  \overline{Q^2}t_aQ^2 -\overline{Q^1}t_aQ^1\, .
\end{equation}
It has been shown that the presence of $S^{\NN=2}_{\textrm{Mag FI},\,\fn}$ shifts the electric $D$ term $D_{\fn}^A$ by an imaginary constant off-shell, allowing us to write
\begin{equation}
 S^{\NN=2}_\textrm{SQCD}+S^{\NN=2}_{\textrm{Mag FI},\,\fn}
=\left.\left(\frac1{16\pi i}\int d^4x\,(D)^4\FF(\WW,\WW^\fn)-\frac1{2}\int\,du\,d\zeta^{-4}\Qt^+\,\DD^{++}\,Q^+\right)\right|_{D^A_\fn\rightarrow\DB^A_\fn}+\hc,
\end{equation}
where
\begin{equation}
\DB_\fn^A=D_\fn^A+4\,i\,\xi_D^A,\qquad \DBb_\fn^A=D_\fn^A-4\,i\,\xib_D^A.
\end{equation}
Taking the full off-shell action as
\begin{equation}
S_\textrm{off-shell}=S^{\NN=2}_\textrm{SQCD}+S^{\NN=2}_{\textrm{El FI},\,\fn}+S^{\NN=2}_{\textrm{Mag FI},\,\fn}
\end{equation}
and solving the $D$ term equations of motion up to third derivatives in the prepotential, we finally arrive at the desired on-shell action for $\NN=2$ SQCD coupled to the ATP mechanism:
\begin{equation}
S_\textrm{on-shell}=\int d^4x\,\left(\LL_{\textrm{kin}}+\LL_{\textrm{yuk}}+\LL_{\textrm{Pauli}}+\LL^\prime_{\textrm{4 Fermi}}+\LL^\prime_{\textrm{D Fermi}}-V^\prime\right),
\end{equation}
where $\LL_{\textrm{kin}}$, $\LL_{\textrm{yuk}}$, and $\LL_{\textrm{Pauli}}$ are unchanged from their respective forms in Appendix \ref{sec:N=2HSS}, and 
\begin{align}
-4\pi\LL^\prime_\textrm{4 Fermi}&= \left(\frac{2\pi i}3\,\FF_{abcd}|\,\lda^a\lda^b-\frac1{8}\,g^{ab}(\FF_{aef}|\,\lda^e\lda^f-\FFb_{afe}|\,\ldab^e\ldab^f)\FF_{bcd}|\right)^A(\lda^c\lda^d)^A+\hc,\\
4\pi\LL^\prime_{\textrm{D Fermi}}&=\frac i2\,\DB^{a,\,\,A}|\;\FF_{abc}|\,(\lda^b\lda^c)^A+\hc\label{eq:4FermiP}\\
4\pi V^\prime &=\frac12\,g_{ab}\,\DB_\phi^{a,\,A}|\;\DBb_\phi^{b,\,A}|+4\pi \overline{Q^i}\{\Xb,X\}\,Q^i-\frac12\,g_{ab}f^a_{cd}f^b_{ef}\Xb^c X^d\Xb^eX^f\nonumber\\ &+4\,i\,(\xi^A+\xib^A)(\xi_D^A-\xib_D^A),\label{eq:scalarpot}
\end{align}
where the solutions of the $D$ term equations of motion have the convenient decomposition
\begin{equation}
\DB^{a,\,A}=\DB^{a,\,A}_X+D^{a,\,A}_Q+D^{a,\,A}_\lambda,\quad \DB^{a,\,A}_\phi=\DB^{a,\,A}_X+D^{a,\,A}_Q,\quad \xiB^{A}_a\equiv(\xi^A+\xib^A)\de^\fn_a+(\xi_D^A+\xib_D^A)\FFb_{\fn a}|
\end{equation}
\begin{equation}
\DB^{a,\,A}_X=-2\,g^{ab}\,
\xiB^{A}_b,
\quad
D^{a,\,A}_Q=-2\,i\,g^{ab}\,\QQ^{A}_b,
\quad
D^{a,\,A}_\lambda=-\frac i2\,g^{ab}\,\FF_{bcd}|(\lda^{c}\lda^{d})^A+\hc.
\end{equation}
We shall refer back to these equations frequently below.

\subsubsection{Evading the \emph{2 into 1 won't go theorem}}

Before we recover the ${\cal N}=1$ lagrangian, it is worth recapping how this mechanism evades the general lore that spontaneous partial SUSY breaking is not possible in flat spacetime. Concretely: it is not possible to begin with an $N$-extended SUSY theory, then, through spontaneous breaking, arrive at an $M$-extended SUSY theory with $N>M>0$. Let us briefly review the argument behind the no-go theorem.
Consider the $N$-extended SUSY algebra
\begin{equation}
\{Q_\al^A,\Qb_{\bed,B}\}=2\,\si_{\al\bed}^\mu\,\de^A{}_B\,P_\mu\,,\qquad A,B=1,\ldots,N.
\label{eq:NextSUSY}
\end{equation}
The vacuum energy in these theories is
\begin{equation}
E_\textrm{vacuum}=\langle 0|H|0\rangle=\frac14\left(||Q^A_1|0\rangle||^2+||\Qb_{1,A}|0\rangle||^2+||Q^A_2|0\rangle||^2+||\Qb_{2,A}|0\rangle||^2\right)
\label{eq:NextSUSYvac}
\end{equation}
and is true for every $A$. This vacuum energy is positive if \emph{any} of the SUSY generators $Q_i^A$ or $\Qb_{i,A}$ do not annihilate the vacuum. It follows if \emph{any} of the SUSY generators are broken, then at least one of $Q_i^A$ or $\Qb_{i,A}$ is broken for \emph{every} $A$ in order for eq. \ref{eq:NextSUSYvac} to hold for all $A$. (The alternative is that none of them are broken.)

The loophole to this argument requires noting that eq. \ref{eq:NextSUSY} isn't always valid in the case of spontaneously broken SUSY. The SUSY algebra in eq. \ref{eq:NextSUSY} follows from the supercurrent algebra
\begin{equation}
\int d^3y\,\{J^A_{\nu,\al}(x),J_{B,0,\bed}(y)\}=2\,\si^\mu_{\al\bed}\,\de^A{}_BT_{\mu\nu}(x).
\label{eq:Supercurrent}
\end{equation}
This is \emph{not} the most general current algebra consistent with SUSY \cite{Hughes1986}, as the Jacobi identities of SUSY \cite{Haag1975} allow an additional field-independent constant piece 
\begin{equation}
\Delta=\si_{\mu,\al\ald}C^A{}_B
\end{equation}
to be added. $\Delta$ commutes with all quantities in the theory so the SUSY algebra on the fields is not modified \cite{Lopuszanski1978}. If $C^A{}_B=0$, then we can integrate eq. \ref{eq:Supercurrent} over the $x$ 3-space to reproduce eq. \ref{eq:NextSUSY} as is usually understood and the no-go theorem holds. When $C^A{}_B\neq0$ there is an infinite contribution to the right hand side of eq. \ref{eq:NextSUSY} from $\Delta\int d^3x$ making the SUSY algebra derived in this manner ill-defined, and allowing evasion of the no-go theorem. The ATP mechanism is precisely a realization of a physical model inducing a non-zero $C^A{}_B$ \cite{Ferrara1996}, where the vacuum energy in the partially broken SUSY vacuum is now related to the FI terms \cite{Fujiwara2004}.

\subsubsection{$\NN=1$ conditions: scalar potential}
\label{sec:scalarpot}
We will now  ensure that the properties of $\NN=2$ SQCD coupled to the ATP mechanism as described in Subsection \ref{sec:breakinginHSS} are those of the $\NN=1$ theory presented in Appendix \ref{sec:2to1N=1SS} as desired. There are three conditions that one could consider for the vacuum to respect $\NN=1$:
\begin{itemize}
\item {\em Vacuum stability}
\item {\em Zero vacuum energy}
\item {\em A scalar potential corresponding to the $\NN=1$ preserving superpotential in eq. \ref{eq:N=1scalardef}.}
\end{itemize}
As we shall see the first two of these provide a constraint on the FI terms while the third is observed to be generally true, and relates the prepotential to the desired $\NN=1$ deformations. In addition, although it is possible to set the vacuum energy to zero, it is not obligatory for preserving $\NN=1$ SUSY \cite{Partouche1997}, but it is natural to apply it. Results for the first two are available in the literature but somewhat scattered, so it is worth collating all three elements here. 

\bigskip

\noindent\underline{\em Vacuum stability}\; : Stable SUSY breaking vacua exist on the Coulomb branch
 (i.e. with $\vev{Q}=0$) which can be achieved by assuming $X^\fn\neq 0$ \cite{Fujiwara2004,Fujiwara2004a,Fujiwara2005,Fujiwara2005a,Fujiwara2006,Itoyama2008} or on the higgs branch when $X^\fn = 0$. 
 In order to study the latter without breaking $SU(N_c)$  one could introduce hypermultiplets charged only under $U(1)_\fn$, but this case is more complicated to analyse as the 
goldstino comes from a linear combination of the new quarks and the $\lambda$'s, so we will restrict the discussion to the former case\footnote{By \emph{Coulomb branch} we are referring to $X^\fn \neq 0$. In this vacuum the hypermultiplets acquire mass from $X^\fn$ but the $X^{\tilde{a}}$ are unconstrained by the 
equations of motion because of the extra degree of freedom provided by $X^\fn$. In the presence of the superpotental term $W\supset X^3$ (assuming that we can eventually make it), setting $X^\fn = 0$ would force some $X^{\tilde{a}}\neq 0 $, with the theory sitting at an Argyres-Douglas point \cite{Argyres1995b}. This would break the gauge symmetry, and may be interesting for phenomenology; we leave this possibility for future study.}. 

Noting that the scalar potential \ref{eq:scalarpot} contains\footnote{We will refer to the scalar potential after $D$ term shifts and substitution as $V$ instead of $V^\prime$ as was used in \ref{eq:scalarpot} in order to avoid confusion with derivatives and to reduce clutter.}
\begin{equation*}
-4\pi V\supset\frac12\,g_{ab}f^a_{cd}f^b_{ef}\Xb^c X^d\Xb^eX^f,
\end{equation*}
it follows that $\vev{X^{\hat{a}}}=0$ where $t_{\hat{a}}$ are non-Cartan generators. Therefore only $\vev{X^\au}\neq0$ is possible, where $t_\au$ are Cartan generators. The vacuum condition is \cite{Fujiwara2004a}
\begin{equation}
4\pi\,\left\langle\frac{\partial V}{\partial (W^a|)}\right\rangle=\frac i4\,\vev{\FF_{abc}\DB^{b,\,A}\DB^{c ,\,A}}=0.\label{eq:vaccum1}
\end{equation}
The only non-vanishing $\vev{\FF_{ab}}$ are the diagonal elements $\vev{\FF_{\hat{a}\,\hat{a}}}$  and $\vev{\FF_{\au\,\au}}$, whilst the only non-vanishing $\vev{\FF_{abc}}$ are $\vev{\FF_{\au\,\au\,\au}}$ and $\vev{\FF_{\au\,\hat{b}\,\hat{b}}}$. 
%(see Appendix of \cite{Fujiwara2005a}). 
It follows that $\vev{\DB^{\hat{a}}}=0$ and so condition \ref{eq:vaccum1} becomes 
\begin{equation}
\vev{\FF_{\au\,\au\,\au}\DB^{\au,\,A}\DB^{\au ,\,A}}=0.
\end{equation}
The choice $\vev{\FF_{\au\,\au\,\au}}=0$ corresponds to unstable saddle points,
and so a stable vacuum must satisfy
\begin{equation}
\langle\DB^{\au,\,A}\DB^{\au,\,A}\rangle=0
\end{equation}
for every $\au$.  By fixing the $SU(2)_R$ direction appropriately, this condition is solved by
\begin{equation}
\label{stab}
\vev{\FF_{\fn\fn}}=-\frac1m\,\left(e+ i\xi\right),\qquad \xi^A+\xib^A=(0,e,\xi)^A,\qquad \xi_D^A+\xib_D^A=(0,m,0)^A,
\end{equation}
where $e,m$ and $\xi$ are real constants. Without loss of generality, taking $\frac\xi m<0$ fixes the sign of the solution as we demand a positive metric, $
 \vev{g_{\fn\,\fn}}=-\frac\xi m\geq0$.
 
\bigskip

\noindent\underline{\em Zero vacuum energy}\; :
The vacuum energy is given by 
\begin{equation}
\vev{4\pi V}=-4\,\xi\,m-4\,i\,(\xi^A+\xib^A)(\xi_D^A-\xib_D^A),
\end{equation}
so that the choice
\begin{equation}
\xi_D^A-\xib_D^A=(0,0,i\,m)^A
\end{equation}
makes it vanish \cite{Antoniadis1996,Itoyama2008}. The form of $\xi_D^A$ is then completely fixed, whereas the imaginary part of $\xi^A$ is still undetermined,
\begin{equation}
\textrm{Re}\,\xi^A=\frac12\,(0,e,\xi)^A,\qquad \xi_D^A=\frac m2\,(0,1,i)^A.
\end{equation}

\bigskip

\noindent\underline{\em A scalar potential corresponding to $W_\textrm{def}$ in \ref{eq:N=1scalardef}}\; : Our third requirement is that we can describe $W_\textrm{def}$ correctly in this setup. The first term in \ref{eq:scalarpot} is
\begin{equation}
\label{eq:scalarpotlast}
4\pi V\supset2\,g^{ab}\,\left[\xiB_a-i\,\QQ_a\right]^A\left[\xiB_b-i\,\QQ_b\right]^{A\,\dagger}.
\end{equation}
From the above, \ref{queues} and \ref{eq:scalarpot}, the $U(1)_\fn$ part of the potential takes the form 
\begin{equation}
V = | X^\fn |^2 |Q^i|^2 +  \frac{g^2}{2}  \left| \overline{Q^2 }Q^1 - \overline{Q^1 }Q^2 \right|^2 + \frac{g^2}{2} \xi^2 + \frac{g^2}{2} \left|\xi - |Q^1|^2 + |Q^2|^2 \right|^2  \, ,
\end{equation}
confirming that it is stable if $X^\fn > g\,\xi$. Note for later reference that along the Coulomb branch the quarks all gain masses and decouple.

Now consider the $SU(N)$ part. The kinetic terms already identify $g_{ab}=\tau_2 K_{ab}$, so in order to reproduce the scalar potential \ref{eq:N=1scalardef}, the above 
together with eq. \ref{implicitqhat} suggest the identification
\begin{equation}
 |{\xiB}_a^{(2)}|\leftrightarrow \frac{4\pi }{\sqrt{2}}\left| \frac{\partial W_\textrm{eff}}{\partial X^a}\right| .
\end{equation}
Defining a rescaled superpotential $\hat{W}_{\mathrm{eff}}={4\pi} W_{\mathrm{eff}}$ (noting that $W^a | = i\sqrt{2} X^a$), this implies
\begin{equation}
 \hat{W}_{\mathrm{eff}}\supset (e\,W^\fn+m\,\FF_\fn)|+\ldots .
\end{equation}
%%%%%%%%%%%%%%%%%%%
Hence a reasonable guess is that in order to preserve an $\NN=1$ SUSY gauge theory with an effective rescaled superpotential $\hat{W}_\textrm{def}$ for the traceless $SU(N)$ adjoint matter (which we will henceforth denote $\Xt$), one should take
\begin{equation}
\label{prep}
\FF(W)=\frac \tau 2\,W^a W^a+\frac{W^\fn}{\Lambda^2}\hat{W}_\mathrm{def},
\end{equation}
where $\Lambda^2 = m$ (which has dimension 2) is the scale of new physics integrated out to form the effective prepotential, and the conditions above give 
Im$(\tau) = -\frac{\xi}{m}$. For example deformations of the Kutasov type can be encoded by simply choosing,
\begin{equation}
\hat{W}_\textrm{def}
\supset 4\pi \frac{\kappa}{k+1}\tr\, \tilde X ^{k+1}.
\end{equation}
Note that in order to reduce clutter, until further notice the $\kappa$ we refer to will be the holomorphic coupling, not the running coupling of the canonically normalised theory. 

%%%%%%%%%%%%%%%%%%%%%%%

Let us check that the $\NN=1$ scalar lagrangian is recovered in the decoupling limit with this prepotential. Sending $e,m,\xi$  to infinity and keeping $\tau$ finite, from eq. \ref{prep} we have
\begin{equation}
%g_{ab}=\begin{pmatrix} \textrm{Im}(\tau)  &  \frac{1}{m}\textrm{Im}(\partial_{\tilde{a}}\hat{W}_{\mathrm{def}}) \\ \frac{1}{m} \textrm{Im}(\partial_{\tilde{b}}\hat{W}_{\mathrm{def}})  & \textrm{Im}(\tau) + 
%\frac{X^\fn}{m}  \textrm{Im}(\partial_{\tilde{a}}\partial_{\tilde{b}}\hat{W}_{\mathrm{def}}) \end{pmatrix} \,;\,  \, 
g^{ab} =\frac{1}{\tau_2^2} \begin{pmatrix} \tau_2 +\frac{1}{m^2\tau_2} \textrm{Im}(\partial_{\tilde{a}}\hat{W}_{\mathrm{def}})^2  &  - \frac{1}{m}\textrm{Im}(\partial_{\tilde{a}}\hat{W}_{\mathrm{def}}) \\ - \frac{1}{m} \textrm{Im}(\partial_{\tilde{b}}\hat{W}_{\mathrm{def}})  & 
\tau_2 + \frac{1}{m^2\tau_2} \textrm{Im}(\partial_{\tilde{a}}\hat{W}_{\mathrm{def}})^2  \end{pmatrix}+ \ldots.
 \end{equation}
After inserting this into eq. \ref{eq:scalarpotlast}, multiple cancellations eventually yield
 \begin{equation}
 4\pi V\supset \frac{2}{\tau_2}  \left| \frac{1}{i\sqrt{2}}\frac{\partial\hat{W}_{\mathrm{def} }}{\partial X^a}  + \QQ_a^{(3)} - i \QQ_a^{(2)} \right|^2 \, .
\end{equation}
Consulting eq. \ref{queues} we see that $\QQ_a^{(3)} - i \QQ_a^{(2)} = {2\pi i} (Q^1-Q^2)(\overline{Q^1} +\overline{Q^2}) $. Therefore the $\NN=1$ superfields
can be identified as  
\begin{equation}
Q\equiv \frac{1}{\sqrt{2} }(Q^1-Q^2) \, \, ; \tilde{Q}\equiv \frac{1}{\sqrt{2} }(\overline{Q^1}+\overline{Q^2})\, , 
\end{equation}
and we find 
 \begin{equation}
 \label{340}
V\supset \frac{4\pi}{\tau_2}  \left| \partial_a {W}_{\mathrm{def} }  + \sqrt{2} Q t_a \tilde{Q} \right|^2.
\end{equation}
This matches the $\NN=1$ expression in eq. \ref{eq:N=1scalardef} and is a non-trivial check of this approach. 

The $U(1)_R$ symmetry of the $\NN=1$ theory is then identified with the $\sigma^1$ generator of $SU(2)_R$, under which $Q$ and $\tilde{Q}$ have the same charge. As discussed above, on the Coulomb branch we have $X^\fn>g\,\xi$ for stability, so the quarks will decouple as well, although one can arrange to keep them in the spectrum by choosing $g_\fn \ll g_{SU(N)}$.

\subsubsection{$\NN=1$ conditions: gaugino-fermion lagrangian}

\label{sec:gaugino-fermion}

A second stringent test is to check that the correct $\NN=1$ fermion lagrangian is also induced by eq. \ref{prep}, as well as the presence of a massless gaugino.

The term providing the fermion contributions coming from the partial SUSY breaking \ref{eq:4FermiP} is
\begin{equation}
4\pi\LL_{\textrm{D Fermi}}=\frac i2\,\DB^{a,\,\,A}|\;\FF_{abc}|\,(\lda^b\lda^c)^A+\hc.\label{eq:fermionneq2}
\end{equation}
This, together with the yukawa interaction
\begin{equation*}
4\pi\LL_\textrm{yuk}\supset\frac i{\sqrt2}\,g_{ab}\,f^b_{cd}\,\lda^{a,i}\,\Xb^c\,\lda_i^d+\hc
\end{equation*}
gives rise to the adjoint fermion masses. Since we are only interested in the phase where $\vev{X^{\hat{a}}}=0$, we can ignore the yukawa term for a spectrum analysis for the $SU(N_c)$ part. For the $U(1)_\fn$ theory this coupling does not exist because there are no abelian self interactions. Noting that $\vev{\FF_{\hat{a}\fn\fn}}=0$, we can decompose \ref{eq:fermionneq2} into the $U(1)_\fn$ and $SU(N_c)$ parts as
\begin{equation}
-\LL_{\textrm{D Fermi}}=\frac12\,M_\fn^{ij}\,\lda_i^\fn\,\lda^\fn_j+\frac12\,M^{ij}\,\lda^\at_i\,\lda^\at_j+\hc
\end{equation}
where the fermion mass matrices are
\begin{equation}
M_\fn^{ij}=\frac{{i}\,g^{\fn\fn}}{4\pi}\begin{pmatrix}e+m\,\FFb_{\fn\fn}| & -i\,\xi\\-i\,\xi & e+m\,\FFb_{\fn\fn}\end{pmatrix}^{ij}\FF_{\fn\fn\fn},\quad M^{ij}=\frac{i\,g^{\fn\fn}}{4\pi}\begin{pmatrix}e+m\,\FFb_{\fn\fn} & -i\,\xi\\-i\,\xi & e+m\,\FFb_{\fn\fn}\end{pmatrix}^{ij}\FF_{\fn\at\at}|.
\end{equation}
In the vacuum determined above \ref{sec:scalarpot} these become
\begin{equation}
M_\fn^{ij}= -\frac{m}{4\pi}\begin{pmatrix}1 & -1\\-1 & 1 \end{pmatrix}^{ij}\vev{\FF_{\fn\fn\fn}},\qquad M^{ij}= -\frac{m}{4\pi}\begin{pmatrix}1 & -1\\-1 & 1 \end{pmatrix}^{ij}\vev{\FF_{\fn\at\at}}.
\end{equation}
Note that the latter term can be rewritten as
\begin{equation}
\label{345}
M^{ij}=  \frac 12
\begin{pmatrix}1 & -1\\-1 & 1 \end{pmatrix}^{ij} 
\frac{ \partial^2{W}_\textrm{def} }{\partial X^{\tilde{a}}\partial X^{\tilde{a}} }.
\end{equation}
This correctly matches eq. \ref{eq:N=1fermiondef} as required.

Since for $m, \vev{\FF_{\fn\fn\fn}},$ and $\vev{\FF_{\fn\at\at}}$ all non-zero we have 
\begin{equation}
\det\,M_\fn\, =\, \det\,M\, =\, 0,\qquad \tr\,M_\fn\neq0,\qquad\tr\,M\neq0,
\end{equation}
the $U(1)_\fn$ fermions and the $SU(N_c)$ fermions each have one linear combination that corresponds to a massless eigenstate, and one linear combination that corresponds to an eigenstate of mass $\frac{m\vev{\FF_{\fn\fn\fn}}}{2\pi}$ and $\frac{m\vev{\FF_{\fn\at\at}}}{2\pi}=\partial_{X^{\tilde{a}}}\partial_{X^{\tilde{b}}}W_{\mathrm{def}}$ respectively. The massless $U(1)_\fn$ combination is the Nambu-Goldstone fermion of partial SUSY breaking, and the massless $SU(N_c)$ combination is the gaugino of the unbroken gauge symmetry as required\footnote{This can be seen by calculating the SUSY transformations where one finds \cite{Fujiwara2004}
\begin{equation}
\vev{\delta_Q\lda^\fn_\textrm{massless}}\sim\vev{\DB^\fn_\textrm{massless}}\neq 0,\qquad \vev{\delta_Q\lda^\at_\textrm{massless}}\sim\vev{\DB^\at_\textrm{massless}}=0.
\end{equation}
}. In the $\NN=1$ preserving vacuum, note that the massless $SU(N_c)$ gaugino does not enter the superpotential, only the (potentially) massive $SU(N_c)$ combination will.

\section{Breaking $\NN=2$ to $\NN=0$ with gaugino masses}

\subsection{Overview}

This requires a further extension of the gauge symmetry to $SU(N_c)\times U(1)^3$; we can then assign a combination of FI terms to pick out an $\NN=1$ preserving direction, and as a perturbation, assign a different combination of FI terms to fully break SUSY. This provides us with a description of an $SU(N_c)$ $\NN=2$ theory augmented by both $\NN=1$ deformations and soft-terms that can, as we shall see in the next section, all be mapped under electric-magnetic duality.

\subsection{A general observation: gaugino masses from additional $U(1)^\prime$s}
\label{sec:breaking2to0}

Having learned how to write  $\NN=1$ SQCD+$X$ theories as the low energy limit of spontaneously broken $\NN=2$ theories, we are in a position to deform the theory further with soft perturbations that arise from the complete spontaneous breaking of SUSY from $\NN=2\rightarrow\NN=0$ by the same mechanism. In the present context we are particularly focussed on Dirac gaugino masses so it is useful to begin with some general observations.

We will be thinking of the additional $U(1)$'s as a perturbation on the $\NN=1$ theory (in the sense that $m_D\ll \Lambda$) and will take the FI-terms for $U(1)_\fn$ to be as described above.  Although Dirac mass-terms can famously preserve an $R$-symmetry, in the context of Kutasov duality they will break it (since the $\NN=1$ gauginos have $R$-charge $1$ and therefore the Dirac mass requires $\tilde{X}$ to have $R$-charge zero, in conflict with $W_{\mathrm{def}}\supset \kappa X^{k+1})$. Therefore the FI-terms for the new $U(1)$'s must have some component along the $\sigma^1$ direction of $SU(2)_R$ which as we saw in section \ref{sec:scalarpot} is the $U(1)_R$ direction of the $\NN=1$ theory. 
Furthermore the contribution from FI-terms to the fermion mass matrix $M^{ij}$ are
$M^{ij} \sim \xiB^A (\si^A \varepsilon)^{ij}$ where $\varepsilon$ is the $SU(2)_R$ metric. But the stability condition essentially fixes $\xiB$ to be null. We can parameterise this generally by taking
$\xiB^A = (\alpha, i \sqrt{\al^2+\be^2}, \be)$ regardless of the origin of $\al$ and $\be$. The stability conditions for $\xiB$ then simply fix the VEVs of the $F_{abc}$ to satisfy this condition (the specific case above has  $\al=0$, $\be=\xi$). Shifting to the basis in which the $\NN=1$ created by $U(1)_\fn$ is diagonal, we find that additional terms from a single extra $U(1)$ are of the form 
\begin{equation}
\label{deltam}
\delta M^{ij}\sim 
\begin{pmatrix}
 -\be + \sqrt{\al^2+\be^2} &  -\al  \\
 -\al                  &  \be + \sqrt{\al^2+\be^2} 
\end{pmatrix} \, .
\end{equation} 
Clearly for any choice of $\al$ and $\be$ one can never set the $\delta M^{11}$ and $\delta M^{22}$ components to zero unless $\al$ is zero as well, and it is therefore impossible to introduce a pure Dirac mass with a single extra $U(1)$. On the other hand it is always possible (by tuning parameters) to do this with two extra $U(1)$'s.

Consider therefore an $SU(N_c)\times U(1)_\fn \times U(1)_\hn \times U(1)_\qn$ theory, where the $Q^+$ is charged under only the $U(1)_\fn$ as displayed in table \ref{tab:n=2fieldsc}.
\begin{table}[ht]
\begin{center}
\begin{tabular}{c|c|c|c|c|c}
             & $SU(N_c)$       & $U(1)_\fn$  & $U(1)_\hn$ & $U(1)_\qn$ & $SU(N_f)$ \vspace{0.06cm}\\ \hline
 $Q^+$ & \tfun               & 1                & 0                 & 0                & \tfun
\end{tabular}
\end{center}
\caption{\em $\NN=2$ superfield representations in $\NN=2$ SQCD coupled to $U(1)_\fn\times U(1)_\hn\times U(1)_\qn$.}
\label{tab:n=2fieldsc}
\end{table}
This theory is in the same form as in \ref{eq:N=2SQCD} and \ref{eq:N=2SQCDparts} with the prepotential $\FF(W,W^\fn,W^\hn,W^\qn)
$ again being a generic function of $\NN=2$ gauge superfields, and the gauge covariant derivative acting on the hypermultiplets
remaining unchanged. The corresponding additional FI-pieces in the action take the same form as in equations \ref{eq:elFI} and \ref{eq:magFI} with the obvious replacement of gauge group. 
As we mentioned in the preamble to this section, the vacuum stability conditions in the $\NN=0$ theory still set
\begin{equation}
\vev{\DB^{\au,\,A}\DB^{\au ,\,A}}=0
\end{equation}
for $\au$'s corresponding to each of the $U(1)$ factors, where as before there is summation over $A$ but not over $\au$.

There are many combinations that one could consider for the prepotential and the new FI-terms. A simple solution is to allow only $\FF_{\fn\hn}$ and $\FF_{\fn\qn}$ mixing, and just electric FI-terms for the $U(1)_\hn$ and $U(1)_\qn$ factors in the $\sigma^1$ and $\sigma^2$ directions (i.e. we are going to add two $\beta=0$ type solutions and make the Majorana masses cancel). The three vacuum stability equations  then translate into the following conditions;
\begin{equation}
\label{conds}
g_{\fn\fn} \re ( \xi^{(2)}_{D,\fn})  = \re ( \xi^{(3)}_{\fn})\,\, ; \,\, 
g_{\fn\hn} \re ( \xi^{(2)}_{D,\fn}) = \re ( \xi^{(1)}_{\hn}) \,\, ; \,\, 
g_{\fn\qn} \re ( \xi^{(2)}_{D,\fn})  = -\re ( \xi^{(1)}_{\qn}) \, .
\end{equation}
The first of these is essentially the same condition as in eq. \ref{stab}. The imaginary parts can be set to satisfy the zero vacuum energy conditions if desired. In order to get non-zero 
gaugino masses the prepotential is of the form 
\begin{equation}
\label{prep2}
\FF(W)=\frac{\tau_{ab}}{2} \,W^a W^b +  \frac{W^\fn}{\Lambda^2}\hat{W}_\mathrm{def} + \frac 1{2\Lambda} ({W^\hn - W^\qn}) W^{\tilde{a}} W^{\tilde{a}}\, ,
\end{equation}
where $\tau_{ab}= \FF_{ab}|$, and we neglect higher order terms in the leading part.
Note that the mass-inducing third term only involves the two additional $U(1)$'s. The contribution to the gaugino masses is of the form 
\begin{equation}
\delta M^{ij} = -\frac{ (\si^A\varepsilon)^{ij}}{4\pi \Lambda} 
\left\{ \xiB_\fn ^A(g^{\fn \hn} - g^{\fn\qn}) + ( g^{\hn \hn} \xiB_\hn^A - g^{\qn\qn}  \xiB_\qn^A ) \right\} \, .
\end{equation}
In order to forbid additional $\NN=1$ mass terms for the adjoints $X^{\tilde{a}}$, we must choose $g^{\fn \hn} = g^{\fn\qn}$ to make the first term vanish. By 
eq. \ref{conds} we then have $\xiB_\hn^{(1)}  = -\xiB_\qn^{(1)} $. Choosing for simplicity 
$g_{\fn\hn} = g_{\fn\qn}\ll g_{\fn\fn},\, g_{\hn\hn} = g_{\qn\qn}$ together with $g_{\hn\qn}=0$, we then have $g_{\fn\hn}=g_{\fn\qn}\equiv -\alpha/m$.
Hence $\xiB_\hn = (\alpha , i \alpha ,0) $ and $\xiB_\qn = (-\alpha , i \alpha ,0) $, giving a gaugino mass matrix of the form 
\begin{equation}
\delta M^{ij} = -\frac{\alpha}{2\pi \Lambda} 
\begin{pmatrix}
 0 &  1 \\
 1 &  0
\end{pmatrix} \, 
\end{equation} 
as required. Along with these terms we expect the super-soft operators of \cite{Fox2002} to be induced in the scalar potential. 
Consulting eq. \ref{eq:scalarpotlast} it is clear that these arise from the
cross terms $g^{\fn\hn} \QQ^\dagger_\hn \xiB_\fn+g^{\fn\qn} \QQ^\dagger_\qn \xiB_\fn+\hc$. 

It is much easier of course to generate pure Majorana masses: it requires only a single additional $U(1)_\hn$, and a prepotential of the form 
\begin{equation}
\label{prepmaj}
\FF(W)=\frac{\tau_{ab}}{2} \,W^a W^b +  \frac{W^\fn}{\Lambda^2}\hat{W}_\mathrm{def} + \frac 1{2\Lambda} {W^\hn } W^{\tilde{a}} W^{\tilde{a}}\, ,
\end{equation}
choosing FI-terms such that $\alpha=0$ in eq. \ref{deltam}. Furthermore, to avoid this becoming 
just another $\NN=1$ mass-term for the adjoint fields, the sign of $\beta$ is chosen so that the non-zero eigenvalue falls in the 
block that has just been identified by the $U(1)_\fn$ FI-terms as belonging to the $\NN=1$ gauginos. That is with 
$\xiB^A_\fn=(0,i\xi,\xi)$ we choose $\xiB^A_\hn=(0,i\beta,-\beta)$, with both $\xi$ and $\beta>0$.

\section{Duality relations for the $\NN=2\rightarrow \NN=0$ theory}
\label{sec:dualrel}
\subsection{$\NN=1$ couplings and gaugino masses}
\label{sec:kutdef}
Let finally return to our objective, which (recall) is to determine how couplings as well as Dirac gaugino masses map under $\NN=2$ duality, and that the prepotential maps consistently under $\NN=2$ duality. 
 We should at this point make clear that we are not about to 
solve the $\NN=2$ system for arbitrary numbers of colours and flavours. Nevertheless it is possible to make general statements about the constraints such a duality should give on the prepotential. This is enough to establish that it contains all the same operators as the weakly coupled electric superpotential. After this use the spurion technique of \cite{Luty1999} determines the precise coefficients. 

The theory can be written in either electric variables
\begin{equation}
W(\phi,\lambda,D,v),\quad \FF
\end{equation}
or dual magnetic ones,
\begin{equation}
W_D(\phi_D,\lambda_D,D_D,v_D),\quad \FF_D\label{eq:mag}\, ,
\end{equation}
with the relations\cite{Ivanov1997}
\begin{equation}
\label{duality-rels}
W_D^a=\frac{\partial\FF}{\partial W_a},\qquad W^a=-\frac{\partial\FF_D}{\partial W_{D,a}}.
\end{equation}
Differentiating this eq. gives the 
functional relation
$
\tau_D=-{\tau}^{-1}\, ,
$
for any prepotential. The mapping of the FI-terms is given by 
\begin{equation}
\xi \rightarrow \xi_D \,\,, \,\,\, \xi_D \rightarrow -\xi\,.
\end{equation} 
Now, it is known that generally the prepotential obeys (in $\NN=1$ language) \cite{Eguchi1995}
\begin{equation}
\label{stones}
\AAcal_{ii}\frac{\partial\FF}{\partial \AAcal_{ii}}-2\FF=8\pi i \beta u
\end{equation} 
where the adjoint modulus $u$ is related to the fields at weak coupling (large $|u|$) as tr$(\AAcal^2)\approx 2u$. If we set $\beta=0$, we find that $\FF=\frac\tau 2 {\AAcal^2} $, so that the RHS of this eq. is encoding the one-loop running of the $\NN=2$ theory. From this eq. we infer 
\begin{equation}
\label{but}
W=W_D /(\tau-8\pi i \beta).
\end{equation}
(The dual version of this eq. has of course a much more complicated $u$ because the theory is strongly coupled.)

Satisfying eq. \ref{stones} for both electric and magnetic theories gives 
\begin{equation} 
\FF_D=\FF-\AAcal_{D,ii} \AAcal_{ii}\,, 
\end{equation} 
so we can also infer that 
\begin{equation} 
\label{beatles}
\FF_D(W_D)=\FF(W(W_{D}))-W_{D} W(W_{D})\,.
\end{equation} 
In other words the magnetic prepotential is given by taking the electric one and replacing $W$ with $W(W_D)$ determined  as a function 
of $W_D$. In general this is extremely complicated, but eq. \ref{but} tells us that $W=W_D /(\tau-8\pi i \beta)$. This is the result we need, because it tells us that, 
while $\tau(W_D)$ will in general be a
complicated function of $W_D$, it is clear that every operator of the electric theory has a direct equivalent in the magnetic theory. 

Indeed, suppose one knows the dual prepotential $\FF_{D}^{(0)} (W_{D})$ of an undeformed $\NN=2$ theory, with prepotential $\FF^{(0)} (W)$. If the 
theory is then deformed to $\FF(W)=\FF^{(0)}+\kappa \FF_\kappa$, where $\kappa$ is parametrically small, then in a $\kappa$ expansion, a dual prepotential of the form \begin{equation}
\FF_D(W_D)=\FF^{(0)}(W^{(0)}(W_D))+\kappa \FF_\kappa(W^{(0)}(W_D)) ,
\end{equation} 
where $W^{(0)}(W_D)$ is the function determined from $W_D=\partial\FF^{(0)}/\partial W$, 
is seen to correctly solve equations \ref{duality-rels} and \ref{beatles} to $\OO (\kappa^2)$. 

Having established this fact, we can utilise the spurion technique of \cite{Luty1999} to fix the coefficients of the terms in the $\kappa$-deformation of the magnetic prepotential. The technique used there extends trivially to give two sets of invariants in the $\NN=2$ theory, namely gaugino mass invariants of the form 
$m_{\mathrm{gaugino}} /g^2 $ and $\kappa$ invariants of the form $\kappa/g^{k+1}$ where we now switch back to $\kappa$ being the physical coupling in the canonically normalised theory. Hence the combination $m_D/g\kappa^{\frac{1}{k+1}}$ is obviously also invariant. Focusing on the Dirac mass, we see that $m_D/g^2$ is an RG invariant but of course only in the $\NN=2$ theory (as in \cite{Luty1999}); away from $\NN=2$, the $h$ and $g$ couplings go their separate ways and $m_D/g^2$ will begin to pick up corrections of order $\kappa^2$, but as we know the combination $m_D/g\kappa^{\frac{1}{k+1}}$ remains an RG invariant even as we flow back to $\NN=1$. 

The dual prepotential required for this mapping to be correct is of the form
\begin{equation}
\label{prep3}
\FF(W_D)= \FF^{(0)}(W_D) +  \frac{W_D^\fn}{\Lambda_D^2}\hat{W}_\mathrm{def}(W_D) + \frac 1{2\Lambda_D} ({W_D^\hn - W_D^\qn}) W_D^{\tilde{a}} W_D^{\tilde{a}}+\OO (\kappa^2 )\, ,
\end{equation}
where $\partial_{W_D}^2\FF_0(W_D)| = -{\tau|}^{-1}$, and the magnetic scale is $\Lambda_D= -(e+i\xi)$.
To check that the latter is correct, we can study the 
dual version of the pre-factor in ${\cal L}_{\mathrm{fermion}}$ of eq. \ref{eq:fermionneq2} which is 
\begin{equation}\label{tmp}
g^{\fn\fn} \xiB^A_\fn F_{\fn\tilde{a}\tilde{a}}= -m (0,i,1) \FF_{\fn\tilde{a}\tilde{a}}.\end{equation}
 In the dual variables we first note that the stability conditions for $\xiB^A_\fn=(0,-m,0) + (0,e,\xi) \bar{\FF}_{D,\fn\fn}$, consistently 
 give ${\FF}_{D,\fn\fn}=m/(e+i\xi)=-1/{\FF}_{\fn\fn}$. Then straightforward manipulation leads to 
 \begin{equation}\tilde{g}^{\fn\fn} \tilde{\xiB}^A_\fn \FF_{D,\fn\tilde{a}\tilde{a}}= ({e+i\xi})  (0,i,1) \FF_{D,\fn\tilde{a}\tilde{a}}.\end{equation}
Note that the magnetic FI-term is usurped by electric ones. Comparison with eq. \ref{tmp} shows that one factor of $\tau$ correctly cancels from the dual coupling, $\Lambda^2_D = {\tau} \Lambda^2$. It is also straightforward to check, although we do not show it explicitly (it is quite a bit more tedious as we need to solve the dual stability conditions with all three FI-terms), that the tuning of FI-terms that gave Dirac masses in the electric theory is the correct tuning for Dirac masses in the magnetic theory -- i.e. we consistently map Dirac to Dirac gaugino, and Majorana to Majorana. 
 
\subsection{Quarks under Electric-Magnetic duality}

Let us briefly comment on the mapping of the quark hypermultiplet $Q^+$ under the $\NN=2$ S-duality. By considering finiteness, the mapping of gauge invariants, and requiring that known non-self dual points are not mapped onto each other, refs.\cite{Leigh1995,Strassler1996} argue that a natural map for $SU(N_c)$ $\NN=2$ SQCD deformed by a mass for the chiral adjoint in the unbroken phase is into a similar theory $SU(N_c)$ $\NN=2$ SQCD\tprime  \, with the charge conjugation acting on the flavour structure. The new hypermultiplets $q^+$ are interepreted as the general $N_c$ case of the semi-classical monopoles of \cite{Seiberg1994c,Seiberg1994b}, and the mass for the chiral adjoint is mapped to itself. For our purposes, we have already shown that a mass for the chiral adjoint is mapped to itself in section \ref{sec:kutdef}, and so we expect the conclusions of \cite{Leigh1995,Strassler1996} to apply here as well.

\section{Conclusions}

We have presented evidence for the invariance of 
\begin{equation}
\frac{ m_{{D}}}{{g} {\kappa}^{\frac{1}{k+1}}}
\end{equation}
under Kutasov duality, where $m_D$ is the Dirac gaugino mass. 
This was achieved by analysing the flow to the $\NN=2$ dual theories in which the $\NN=1$ deformations and gaugino masses were generated by FI-terms of additional $U(1)$ factors coupling in deformations in the prepotential. Such a flow was found to occur in 
the perturbative case, and it was also found that the ${\cal N}=2$ Yukawa induces the correct higgsing in the dual theory, even in the non-perturbative case. 

Along the way, we discussed the generalities of embedding $\NN=1$ terms within manifestly $\NN=2$ supersymmetric theories 
using the techniques of harmonic superspace. Although the formalism is somewhat cumbersome, it has the advantage that quarks can be treated appropriately off-shell, $SU(2)_R$ symmetry breaking is made manifest and dynamical, and the interplay between $\NN=0$ terms (i.e. gaugino masses) and $\NN=1$ terms is evident. Aside from its obvious direct application to Dirac gaugino phenomenology, our results could therefore be useful for constructing an entirely dynamical realisation of $\NN=2$ sectors within an $\NN=1$ theory, as has often been proposed for the higgs and gauge sectors (see \cite{Fox2002} and more recently \cite{Heikinheimo2011}).

\subsection*{Acknowledgements}
We thank Valya Khoze, Jan Louis, Veronica Sanz and Paul Smyth for support and discussions. DB is supported by a U.K. Science and Technology Facilities Council (STFC) studentship.
 
\newpage 
 
\appendix

\section{$\NN=2$ SQCD}

\subsection{Total effective lagrangian}

\label{sec:N=2HSS}

The lagrangian $\LL_\textrm{total}$ for $\NN=2$ SQCD arising from eqs. \ref{eq:N=2SQCD}, \ref{eq:N=2SQCDparts}, and \ref{eq:prepotential} up to four derivatives in the prepotential $\FF(\WW)$ is 
\begin{equation}
\LL_\textrm{total}=\LL_\textrm{kin}+\LL_\textrm{yuk}+\LL_{\textrm{Pauli}}+\LL_\textrm{D Fermi}+\LL_\textrm{4 Fermi}-V
\end{equation}
where
\begin{align}
-\LL_\textrm{kin}&=\frac{g_{ab}}{4\pi}\left(\DD^\mu X^a\,\DD_\mu\Xb^b+i\,\lda^{i,\,a}\,\si^\mu\,\DD_\mu\,\ldab^b_i-\frac14\,F_{\mu\nu}^a\,F^{b,\,\mu\nu}\right)+\frac{h_{ab}}{16\pi}\,F_{\mu\nu}^a\,\tilde{F}^{b,\,\mu\nu}\nonumber\\
&+\Qb^i \,\DD^\mu\DD_\mu \,Q_i+\frac i2\left(\psib_Q\sib^\mu\DD_\mu\psi_Q+\psi_\Qt\si^\mu\DD_\mu\psib_\Qt\right),\label{eq:lkin} \\
-\LL_\textrm{yuk}&=\,\frac{i g_{ab}}{ 4\pi {\sqrt2}}\,f^b_{cd}\,\lda^{a,i}\,\Xb^c\,\lda_i^d+i\left(\Qb^i\,\lda_i\,\psi_Q-\psi_\Qt\,\lda^i\,Q_i\right)-\frac1{\sqrt2}\,\psi_\Qt\,X\,\psi_Q+\hc,\label{eq:lyuk}\\
\LL_\textrm{D Fermi}&=\frac i2\,\FF_{abc}|(\lda^a\lda^b)^AD^{c,A}+\hc, \\
-\LL_\textrm{Pauli}&= \frac i4\,\FF_{abc}|\,\lda^{a,i}\si^{\mu\nu}\lda^b_i\,F_{\mu\nu}^c+\hc, \\
-\LL_\textrm{4 Fermi}&= \frac i6\,\FF_{abcd}|\,(\lda^a\lda^b)^A(\lda^c\lda^d)^A+\hc,\\
V&=\Qb^i\,\{\Xb,X\}\,Q_i-\frac{g_{ab}}{4\pi} \left(\frac12\,f^a_{cd}\,f^b_{ef}\Xb^c X^d\Xb^e X^f+\frac12\,D^{a,\,A}|\; D^{b,\,A}|\,\right),
\end{align}
an traced $SU(2)_R$ tensor products are written as three vector dot products
\begin{equation}
a^i{}_j\equiv i\,a^A\left(\si^A\right)^i{}_j,\qquad a^{ij}\,b_{ij}=-a^i{}_j\,b^j{}_i=a^A\,b^B\,\tr_R\,(\si^A\cdot\si^B)=2\,a^A\,b^A,
\end{equation}
where $\tr_R$ is a trace over the $SU(2)_R$ indices and we use the conventions of Appendix \ref{sec:indexandsu2r}. The standard renormalizable $\NN=2$ SQCD lagrangian can be obtained by integrating out the $D^{a,A}$ and taking the canonical prepotential
\begin{equation}
\FF(W)=\tau \frac{(W^a)^2}2,
\qquad \tau
\equiv \frac{\tta_\textrm{YM}}{2\pi}+\frac{4\pi i}{g^2}
\equiv \tau_1+i\,\tau_2,\qquad \tau_1,\,\tau_2\in\RRRR
\label{eq:canpre}.
\end{equation}
One then finds the kinetic terms in the holomorphic basis
\begin{align}
-\LL_\textrm{kin}&=\frac1{g^2}\left(\DD^\mu X^a\,\DD_\mu\Xb_a+i\,\lda^{i,\,a}\,\si^\mu\,\DD_\mu\,\ldab_{i,\,a}+\frac14\,F_{\mu\nu}^a\,F^{\mu\nu}_a\right)
+\frac{\tta_\textrm{YM}}{32\pi^2}\,F^a_{\mu\nu}\,\tilde{F}_a^{\mu\nu},
\end{align}
as well as the familiar yukawa interactions and scalar potential.

\subsection{Formulation in $\NN=1$ superspace}

\label{sec:N=2SQCD,N=1SS}

Because we are ultimately interested in $\NN=1$ SQCD+$X$, we briefly recall how $\NN=2$ SQCD can be recast in $\NN=1$ superspace \cite{Salam1975}. 
\begin{table}[ht]
\begin{center}
\begin{tabular}{c|c|c|c}
 & $SU(N_c)$ & $SU(N_f)$ & $U(1)_R$ \\ \hline
$Q$ & \tfun & \tfun & $1-R_X\frac{N_c}{N_f}$ \\
$\Qt$ & \tafun & \tafun & $1-R_X\frac{N_c}{N_f}$\\
$X$ & \tAd & \ttriv & $R_X$
\end{tabular}
\end{center}
\caption{\em $\NN=1$ superfield representations in $\NN=2$ SQCD.}
\label{tab:n=1fields}
\end{table}
The appropriate $\NN=1$ superfield content \cite{Seiberg1994b,Argyres1996}
is given in table \ref{tab:n=1fields}, and the $\NN=2$ SQCD action composed of two parts as in \ref{eq:N=2SQCD}. From the full $\NN=2$ superspace point of view, after fixing an $SU(2)_R$ direction so that $\ScrQ_1$ is the canonical $\NN=1$ SUSY, the field content of $Q^+$ and $W$ is most easily seen diagramatically in component superfield `diamonds' \cite{Fayet1986},
\begin{figure}[ht]
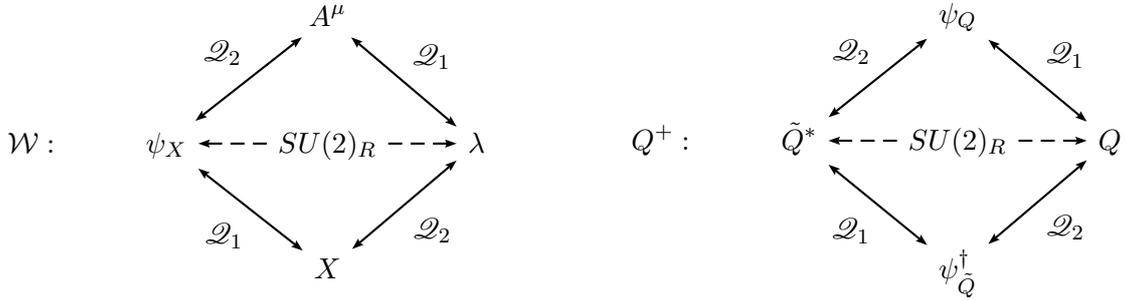

\begin{equation*}
  \psmatrix[colsep=1.2cm,rowsep=1.2cm,nodesep=0.15cm] 
  &&A^\mu&\\ 
  \WW:&\psi_X&SU(2)_R&\lambda\\
  &&X&
  \psset{arrowscale=1}
  \psset{arrowinset=0.15}
  \ncline{<->}{1,3}{2,2}\nbput{\ScrQ_2}
  \ncline{<->}{1,3}{2,4}\naput{\ScrQ_1}
  \ncline{<->}{3,3}{2,2}\naput{\ScrQ_1}
  \ncline{<->}{3,3}{2,4}\nbput{\ScrQ_2}
   \ncline[linestyle=dashed]{<-}{2,2}{2,3}
  \ncline[linestyle=dashed]{->}{2,3}{2,4}
  \endpsmatrix
    \qquad \quad \qquad
  \psmatrix[colsep=1.2cm,rowsep=1.2cm,nodesep=0.15cm] 
  &&\psi_Q&\\ 
  Q^+:&\Qt^*&SU(2)_R&Q\\
  &&\psi_\Qt^\dagger&
  \psset{arrowscale=1}
  \psset{arrowinset=0.15}
  \ncline{<->}{1,3}{2,2}\nbput{\ScrQ_2}
    \ncline{<->}{1,3}{2,4}\naput{\ScrQ_1}
  \ncline{<->}{3,3}{2,2}\naput{\ScrQ_1}
  \ncline{<->}{3,3}{2,4}\nbput{\ScrQ_2}
  \ncline[linestyle=dashed]{<-}{2,2}{2,3}
  \ncline[linestyle=dashed]{->}{2,3}{2,4}
  \endpsmatrix
\end{equation*}
\caption{\em $\mathcal N=2$ SQCD particle content}
\label{fig:n=2particle}
\end{figure}

The SYM part is written in terms of an analytic prepotential $\FF(i\sqrt{2} X)= \FF(\AAcal)$ \cite{Seiberg1988},
\begin{equation}
S^{\NN=2}_\textrm{SYM}=
\frac1{16\pi i}\int d^4x\,d^2\tta\,\left(\,\FF_{ab}\WW^a\WW^b - \int d^2\ttab\, \frac{i}{\sqrt{2}} \FF_a(e^V)^a{}_b\Xb^b\right)+\hc
\end{equation}
whereas the QCD part is
\begin{equation}S^{\NN=2}_Q= \int d^4x\, d^2\tta\left(\sqrt2\,\Qt\,X\,Q + \frac{1}{2}\int d^2\ttab\left[K_Q+K_\Qt\right]\right)\,+\hc
\end{equation}
and $K_\phi$ is the K\"ahler potentials for the superfield $\phi$.  The 
K\"ahler potential for $X$ and effective gauge coupling for the standard renormalisable $\NN=2$ theory can be recovered by taking \ref{eq:canpre},
\begin{equation}
\FF(\AAcal)=\tau\frac{(\AAcal^a)^2}2\implies S^{\NN=2}_\textrm{SYM}=
\frac{\tau}{4\pi i}
\int d^4x\,d^2\tta\,\left(\frac14\,\WW^2 +\frac12\int d^2\ttab\,  K_X\right)+\hc
\end{equation}

\section{$\NN=2$ SQCD in the presence of $\NN=1$ preserving $W(X)$}

\subsection{Formulation in $\NN=1$ superspace}

\label{sec:2to1N=1SS}

Before reviewing the ATP mechanism in HSS, we will first derive the lagrangian for $\NN=2$ SQCD in the presence of a superpotential $W_\textrm{def}(X)$ in $\NN=1$ superspace. For convenience, the $\NN=1$ superspace formulation of $\NN=2$ SQCD is presented in Appendix \ref{sec:N=2SQCD,N=1SS}. Concretely, an $\NN=2$ breaking $X$ deformation $W_\textrm{def}(X)$ causes the shift in in the action
\begin{equation}
S^{\NN=2}_\textrm{SQCD}\rightarrow S^{\NN=2}_\textrm{SQCD}+S^{\NN=1}_\textrm{def},\qquad S^{\NN=1}_\textrm{def}=
  \int d^4x\,d^2\tta\,W_\textrm{def}(X)+\hc
\end{equation}
and yields the additional terms in the lagrangian
\begin{align}
  V_\textrm{def}&=  \frac{4\pi}{\tau_2} K_X^{ab} \left[\frac{\partial W_\textrm{def}}{\partial X^a}+\sqrt2\,\Qt\, t_a\,Q\right]\left[\frac{\partial W_\textrm{def}}{\partial X^b}+\sqrt2\,\Qt\, t_b\, Q\right]^\dagger\label{eq:N=1scalardef}\\
\LL^\textrm{fermion}_\textrm{def}&=-\frac{1}{2}\frac{\partial^2W_\textrm{def}}{\partial X^a\partial X^b}\,\psi^a_X\,\psi^b_X+\hc,\label{eq:N=1fermiondef}
\end{align}
where $K_X^{ab}$ is the inverse of the K\"ahler metric for the physically normalised $X$
\begin{equation}
(K_X)_{ab}\equiv\frac{\partial^2K_X}{\partial X^a\partial\Xb^b},
\end{equation}
and $\tau_2=\tfrac{4\pi}{g^2}$ is the imaginary part of the holomorphic gauge coupling defined in
eq. \ref{eq:canpre}.

\section{Index and $SU(2)_R$ conventions}
\label{sec:indexandsu2r}
The index conventions used can be found in table \ref{tab:indices}.
\begin{table}[h]
\begin{center}
\begin{tabular}{ccc}
Label & Type & Range\\ \hline
$\mu,\nu,\rho,\sigma$ & space-time & $0\;\textrm{to}\;3$\\
$\al,\ald,\be,\bed$ & spinor & $1,2$\\
$i,j,k,l$ & $SU(2)$ & $1,2$\\
$\at,\bt,\ct,\tilde d$ & $SU(N_c)$ adjoint & $1\;\textrm{to}\;(N_c^2-1)$\\
$a,b,c,d$ & all adjoints & $\fn,\hn,\qn,\;1\;\textrm{to}\;(N_c^2-1)$
\end{tabular}
\end{center}
\caption{Index conventions used throughout unless otherwise stated.}
\label{tab:indices}
\end{table}
Our $SU(2)_R$ conventions are $
\ep^{12}=+1$,
and that if
$
a^i{}_j\equiv i\,a^A(\si^A)^i{}_j
$
then clearly
$
a^A=\frac{1}{2i}\tr(\si^A\,a),
$
and in components
\begin{equation}
a^i{}_j=\begin{pmatrix}i\,a^3 & i\,a^1+a^2\\i\,a^1-a^2& -i\,a^3\end{pmatrix},\quad a^{ij}=\begin{pmatrix} i\,a^1 +a^2 & -i\,a^3\\ -i\,a^3 & -i\,a^1+a^2\end{pmatrix},\quad a_{ij}=\begin{pmatrix} -i\,a^1 +a^2 & i\,a^3\\ i\,a^3 & i\,a^1+a^2\end{pmatrix}.
\end{equation}

\section{HSS notation}
\label{sec:hhsnotation}

\subsection{Conjugation rules}
Complex conjugation $\bar\OO$ is defined as
\begin{align}
\overline{\tta_{\al i}}&=\bar\tta^i_\ald,& \overline{\tta_\al^i}&=-\bar\tta_{\ald i};\\
\overline{u^{+i}}&=u_i^-,& \overline{u_i^+}&=-u^{-i};\\
\overline{f^{i_1\ldots i_n}}&\equiv\bar f_{i_1\ldots i_n}, & \overline{f_{i_1\ldots i_n}}&=(-1)^n\bar f^{i_1\ldots i_n}.
\end{align}
Antipodal conjugation $\OO^\star$
\begin{align}
(u^{+i})^\star&=u^{-i},& (u^+_i)^\star&=u^-_i,\\
(u^{-i})^\star&=-u^{+i},& (u^-_i)^\star&=-u^+_i.
\end{align}
Combined complex and antipodal conjugation $(\bar\OO)^\star=\overline{(\OO)^\star}\equiv\widetilde\OO$
\begin{align}
\widetilde{(u_i^\pm)}=u^{\pm i}, & & \widetilde{(u^{\pm i})}=-u^\pm_i.
\end{align}
It is convenient to note that
\begin{equation}
\overline{Q^1}=\Qb_1=\ep_{12}\,\Qb^2=-\Qb^2=\overline{Q_2},\qquad \overline{Q^2}=\Qb_2=\ep_{21}\,\Qb^1=\Qb^1=-\overline{Q_1}.
\end{equation}

% \newpage 

\subsection{Basis and measures}

The harmonic analytic basis is defined as
\begin{equation}
x_A^\mu\equiv x^\mu-2i\tta^{(i}\si^\mu\ttab^{j)}u_i^+u_j^-,\qquad (x_A,\tta^+\ttab^+,u_i^\pm)\equiv (\zeta,u)\, .
\end{equation}
The covariant derivative and scalar projection are
\begin{equation}
\DD^{++}=D^{++}+iV^{++},\qquad
\qquad \OO|\equiv\OO|_{\tta^\pm=\ttab^\pm=0}.
\end{equation}
The measures are defined as
\begin{align}
\int du\,d^{12}X&\equiv\int du\,d^4x\,d^8\tta=\int du\,d^4x_A\,d^4\tta^+d^4\tta^-=\frac1{256}\int du\,d^4x_A\,(D^-)^2(\Db^-)^2(D^+)^2(\Db^+)^2,\\
\int du\,d\zeta^{(-4)}&\equiv\int du\,d^4x_A\,d^4\tta^+=\frac1{16}\int du\,d^4x_A\,(D^-)^2(\Db^-)^2,
\end{align}
with normalisations
\begin{equation}
\int d^8\tta\,\tta^8=\int d^4\tta^+\,(\tta^+)^4=\int d^4\tta\,(\tta)^4=\int d^4\ttab\,(\ttab)^4=1.
\end{equation}
where
\begin{align}
\tta^8&=(\tta^+)^4(\tta^-)^4=(\tta)^4(\ttab)^4,& (\tta^\pm)^4&=(\tta^\pm)^2(\ttab^\pm)^2,\\
(\tta)^4&=(\tta^+)^2(\tta^-)^2,& (\ttab)^4&=(\ttab^+)^2(\ttab^-)^2.
\end{align}

\newpage

\bibliographystyle{JHEP}
\bibliography{mapping_dg_masses}

\providecommand{\href}[2]{#2}\begingroup\raggedright\begin{thebibliography}{10}

\bibitem{Fayet1978}
P.~Fayet, {\it {Massive gluinos}},  {\em Physics Letters B} {\bf 78} (1978),
  no.~4 417--420.

\bibitem{Polchinski1982}
J.~Polchinski and L.~Susskind, {\it {Breaking of supersymmetry at intermediate
  energy}},  {\em Physical Review D} {\bf 26} (1982), no.~12 3661--3673.

\bibitem{Hall1991}
L.~J. Hall and L.~Randall, {\it {U(1)\_R symmetric supersymmetry}},  {\em
  Nuclear Physics B} {\bf 352} (1991) 289--308.

\bibitem{Fox2002}
P.~J. Fox, A.~E. Nelson, and N.~Weiner, {\it {Dirac Gaugino Masses and
  Supersoft Supersymmetry Breaking}},  {\em Journal of High Energy Physics}
  {\bf 2002} (Aug., 2002) 035--035,
  [\href{http://xxx.lanl.gov/abs/0206096}{{\tt 0206096}}].

\bibitem{Nelson2002a}
A.~E. Nelson, N.~Rius, V.~Sanz, and M.~Unsal, {\it {The MSSM without mu term}},
   \href{http://xxx.lanl.gov/abs/0211102}{{\tt 0211102}}.

\bibitem{Antoniadis2005}
I.~Antoniadis, A.~Delgado, K.~Benakli, M.~Quir\'{o}s, and M.~Tuckmantel, {\it
  {Splitting extended supersymmetry}},  {\em Physics Letters B} {\bf 634}
  (Mar., 2006) 302--306, [\href{http://xxx.lanl.gov/abs/0507192}{{\tt
  0507192}}].

\bibitem{Antoniadis2006}
I.~Antoniadis, K.~Benakli, A.~Delgado, and M.~Quiros, {\it {A New Gauge
  Mediation Theory}},  \href{http://xxx.lanl.gov/abs/0610265}{{\tt 0610265}}.

\bibitem{Antoniadis2006a}
I.~Antoniadis, K.~Benakli, A.~Delgado, M.~Quir\'{o}s, and M.~Tuckmantel, {\it
  {Split extended supersymmetry from intersecting branes}},  {\em Nuclear
  Physics B} {\bf 744} (June, 2006) 156--179,
  [\href{http://xxx.lanl.gov/abs/0601003}{{\tt 0601003}}].

\bibitem{Hsieh2008}
K.~Hsieh, {\it {Pseudo-Dirac bino dark matter}},  {\em Physical Review D} {\bf
  77} (Jan., 2008) 015004, [\href{http://xxx.lanl.gov/abs/0708.3970}{{\tt
  arXiv:0708.3970}}].

\bibitem{Amigo2008}
S.~D.~L. Amigo, A.~E. Blechman, P.~J. Fox, and E.~Poppitz, {\it {R -symmetric
  gauge mediation}},  {\em Journal of High Energy Physics} {\bf 2009} (Jan.,
  2009) 018--018, [\href{http://xxx.lanl.gov/abs/0809.1112}{{\tt
  arXiv:0809.1112}}].

\bibitem{Choi2008a}
S.~Choi, M.~Drees, A.~Freitas, and P.~Zerwas, {\it {Testing the Majorana nature
  of gluinos and neutralinos}},  {\em Physical Review D} {\bf 78} (Nov., 2008)
  095007, [\href{http://xxx.lanl.gov/abs/0808.2410}{{\tt arXiv:0808.2410}}].

\bibitem{Choi2008}
S.~Choi, M.~Drees, J.~Kalinowski, J.~Kim, E.~Popenda, and P.~Zerwas, {\it
  {Color-octet scalars of supersymmetry at the LHC}},  {\em Physics Letters B}
  {\bf 672} (Feb., 2009) 246--252,
  [\href{http://xxx.lanl.gov/abs/0812.3586}{{\tt arXiv:0812.3586}}].

\bibitem{Blechman2009}
A.~E. Blechman, {\it {R-symmetric Gauge Mediation and the MRSSM}},  {\em Modern
  Physics Letters A} {\bf 24} (Mar., 2009) 14,
  [\href{http://xxx.lanl.gov/abs/0903.2822}{{\tt arXiv:0903.2822}}].

\bibitem{Benakli2008}
K.~Benakli and M.~D. Goodsell, {\it {Dirac gauginos in general gauge
  mediation}},  {\em Nuclear Physics B} {\bf 816} (July, 2009) 185--203,
  [\href{http://xxx.lanl.gov/abs/0811.4409}{{\tt arXiv:0811.4409}}].

\bibitem{Belanger2009}
G.~Belanger, K.~Benakli, M.~D. Goodsell, C.~Moura, and A.~Pukhov, {\it {Dark
  matter with Dirac and Majorana gaugino masses}},  {\em Journal of Cosmology
  and Astroparticle Physics} {\bf 2009} (Aug., 2009) 027--027,
  [\href{http://xxx.lanl.gov/abs/0905.1043}{{\tt arXiv:0905.1043}}].

\bibitem{Choi2009}
S.~Choi, J.~Kalinowski, J.~Kim, and E.~Popenda, {\it {Scalar gluons and Dirac
  gluinos at the LHC}},  \href{http://xxx.lanl.gov/abs/0911.1951}{{\tt
  arXiv:0911.1951}}.

\bibitem{Benakli2010a}
K.~Benakli and M.~D. Goodsell, {\it {Dirac gauginos and kinetic mixing}},  {\em
  Nuclear Physics B} {\bf 830} (Mar., 2010) 315--329,
  [\href{http://xxx.lanl.gov/abs/1003.4957}{{\tt arXiv:1003.4957}}].

\bibitem{Chun2010}
E.~J. Chun, J.-C. Park, and S.~Scopel, {\it {Dirac gaugino as leptophilic dark
  matter}},  {\em Journal of Cosmology and Astroparticle Physics} {\bf 2010}
  (Feb., 2010) 015--015, [\href{http://xxx.lanl.gov/abs/0911.5273}{{\tt
  arXiv:0911.5273}}].

\bibitem{Benakli2010}
K.~Benakli and M.~D. Goodsell, {\it {Dirac Gauginos, Gauge Mediation and
  Unification}},  \href{http://xxx.lanl.gov/abs/1003.4957}{{\tt
  arXiv:1003.4957}}.

\bibitem{Carpenter2010}
L.~M. Carpenter, {\it {Dirac Gauginos, Negative Supertraces and Gauge
  Mediation}},  \href{http://xxx.lanl.gov/abs/1007.0017}{{\tt
  arXiv:1007.0017}}.

\bibitem{Kribs2010}
G.~D. Kribs, T.~Okui, and T.~Roy, {\it {Viable gravity-mediated supersymmetry
  breaking}},  {\em Physical Review D} (2010).

\bibitem{Choi2010}
S.~Choi, D.~Choudhury, A.~Freitas, J.~Kalinowski, J.~Kim, and P.~M. Zerwas,
  {\it {Dirac neutralinos and electroweak scalar bosons of N = 1/N = 2 hybrid
  supersymmetry at colliders}},  {\em Journal of High Energy Physics} {\bf
  2010} (Aug., 2010) 25, [\href{http://xxx.lanl.gov/abs/1005.0818}{{\tt
  arXiv:1005.0818}}].

\bibitem{Abel2011a}
S.~Abel and M.~D. Goodsell, {\it {Easy Dirac gauginos}},  {\em Journal of High
  Energy Physics} {\bf 2011} (June, 2011) 64,
  [\href{http://xxx.lanl.gov/abs/1102.0014}{{\tt arXiv:1102.0014}}].

\bibitem{Davies2011a}
R.~Davies, J.~March-Russell, and M.~McCullough, {\it {A supersymmetric one
  Higgs doublet model}},  {\em Journal of High Energy Physics} {\bf 2011}
  (Apr., 2011) 108, [\href{http://xxx.lanl.gov/abs/1103.1647}{{\tt
  arXiv:1103.1647}}].

\bibitem{Benakli2011}
K.~Benakli, M.~D. Goodsell, and A.-K. Maier, {\it {Generating $\mu$ and B$\mu$
  in models with Dirac gauginos}},  {\em Nuclear Physics B} {\bf 851} (Oct.,
  2011) 445--461.

\bibitem{Benakli2011a}
K.~Benakli, {\it {Dirac Gauginos: A User Manual}},  {\em ArXiv} (June, 2011) 4,
  [\href{http://xxx.lanl.gov/abs/1106.1649}{{\tt arXiv:1106.1649}}].

\bibitem{Heikinheimo2011}
M.~Heikinheimo, M.~Kellerstein, and V.~Sanz, {\it {How Many Supersymmetries?}},
   {\em Simulation} (Nov., 2011) 7,
  [\href{http://xxx.lanl.gov/abs/1111.4322}{{\tt arXiv:1111.4322}}].

\bibitem{Itoyama2011}
H.~Itoyama and N.~Maru, {\it {D-term Dynamical Supersymmetry Breaking
  Generating Split N=2 Gaugino Masses of Mixed Majorana-Dirac Type}},
  \href{http://xxx.lanl.gov/abs/1109.2276}{{\tt arXiv:1109.2276}}.

\bibitem{Kribs2012}
G.~D. Kribs and A.~Martin, {\it {Supersoft supersymmetry is super-safe}},  {\em
  Physical Review D} {\bf 85} (June, 2012) 115014,
  [\href{http://xxx.lanl.gov/abs/1203.4821}{{\tt arXiv:1203.4821}}].

\bibitem{Davies2012}
R.~Davies, {\it {Dirac gauginos and unification in F-theory}},  {\em Journal of
  High Energy Physics} {\bf 2012} (Oct., 2012) 10,
  [\href{http://xxx.lanl.gov/abs/1205.1942}{{\tt arXiv:1205.1942}}].

\bibitem{Goodsell2012}
M.~D. Goodsell, {\it {Two-loop RGEs with Dirac gaugino masses}},
  \href{http://xxx.lanl.gov/abs/1206.6697}{{\tt arXiv:1206.6697}}.

\bibitem{Benakli2012}
K.~Benakli, M.~D. Goodsell, and F.~Staub, {\it {Dirac Gauginos and the 125 GeV
  Higgs}},  \href{http://xxx.lanl.gov/abs/1211.0552}{{\tt arXiv:1211.0552}}.

\bibitem{Itoyama2013}
H.~Itoyama and N.~Maru, {\it {D-term Triggered Dynamical Supersymmetry
  Breaking}},  \href{http://xxx.lanl.gov/abs/1301.7548}{{\tt arXiv:1301.7548}}.

\bibitem{Marti2001}
D.~Mart\'{\i} and A.~Pomarol, {\it {Supersymmetric theories with compact extra
  dimensions in N=1 superfields}},  {\em Physical Review D} {\bf 64} (Oct.,
  2001) 105025, [\href{http://xxx.lanl.gov/abs/0106256}{{\tt 0106256}}].

\bibitem{Abel2010a}
S.~Abel and T.~Gherghetta, {\it {A slice of AdS 5 as the large N limit of
  Seiberg duality}},  {\em Journal of High Energy Physics} {\bf 2010} (Dec.,
  2010) 91, [\href{http://xxx.lanl.gov/abs/1010.5655}{{\tt arXiv:1010.5655}}].

\bibitem{Argurio2012}
R.~Argurio, M.~Bertolini, L.~{Di Pietro}, F.~Porri, and D.~Redigolo, {\it
  {Holographic Correlators for General Gauge Mediation}},
  \href{http://xxx.lanl.gov/abs/1205.4709}{{\tt arXiv:1205.4709}}.

\bibitem{Kutasov1995}
D.~Kutasov, {\it {A comment on duality in N = 1 supersymmetric non-abelian
  gauge theories}},  {\em Physics Letters B} {\bf 351} (May, 1995) 230--234,
  [\href{http://xxx.lanl.gov/abs/9503086}{{\tt 9503086}}].

\bibitem{Kutasov1996}
D.~Kutasov, A.~Schwimmer, and N.~Seiberg, {\it {Chiral rings, singularity
  theory and electric-magnetic duality}},  {\em Nuclear Physics B} {\bf 459}
  (1996), no.~3 455--496.

\bibitem{Brodie1998}
J.~H. Brodie and M.~J. Atrassler, {\it {Patterns of duality in N = 1 SUSY gauge
  theories}},  {\em Nuclear Physics B} {\bf 524} (July, 1998) 224--250,
  [\href{http://xxx.lanl.gov/abs/9611197}{{\tt 9611197}}].

\bibitem{Intriligator1995b}
K.~Intriligator, R.~Leigh, and M.~J. Strassler, {\it {New examples of duality
  in chiral and non-chiral supersymmetric gauge theories}},  {\em Nuclear
  Physics B} {\bf 456} (Dec., 1995) 567--621,
  [\href{http://xxx.lanl.gov/abs/9506148}{{\tt 9506148}}].

\bibitem{Karch1998}
A.~Karch, T.~Kobayashi, J.~Kubo, and G.~Zoupanos, {\it {Infrared behaviour of
  softly broken SQCD and its dual}},  {\em Physics Letters B} {\bf 441} (Nov.,
  1998) 235--242, [\href{http://xxx.lanl.gov/abs/9808178}{{\tt 9808178}}].

\bibitem{Evans1995}
N.~Evans, S.~D. HSU, M.~Schwetz, and S.~B. Selipsky, {\it {Exact results and
  soft breaking masses in supersymmetric gauge theory}},  {\em Nuclear Physics
  B} {\bf 456} (Dec., 1995) 205--218,
  [\href{http://xxx.lanl.gov/abs/9508002}{{\tt 9508002}}].

\bibitem{Aharony1995a}
O.~Aharony, J.~Sonnenschein, M.~E. Peskin, and S.~Yankielowicz, {\it {Exotic
  nonsupersymmetric gauge dynamics from supersymmetric QCD}},  {\em Physical
  Review D} {\bf 52} (Nov., 1995) 6157--6174,
  [\href{http://xxx.lanl.gov/abs/9507013}{{\tt 9507013}}].

\bibitem{Evans1995a}
N.~Evans, S.~D. Hsu, and M.~Schwetz, {\it {Exact results in softly broken
  supersymmetric models}},  {\em Physics Letters B} {\bf 355} (Aug., 1995)
  475--480, [\href{http://xxx.lanl.gov/abs/9503186}{{\tt 9503186}}].

\bibitem{Giudice1998a}
G.~F. Giudice and R.~Rattazzi, {\it {Extracting supersymmetry-breaking effects
  from wave-function renormalization}},  {\em Nuclear Physics B} {\bf 511}
  (1998), no.~1-2 25--44.

\bibitem{Arkani-Hamed1998}
N.~Arkani-Hamed, G.~F. Giudice, M.~A. Luty, and R.~Rattazzi, {\it
  {Supersymmetry-breaking loops from analytic continuation into superspace}},
  {\em Physical Review D} {\bf 58} (Oct., 1998) 115005,
  [\href{http://xxx.lanl.gov/abs/9803290}{{\tt 9803290}}].

\bibitem{Cheng1998}
H.-c. Cheng and Y.~Shadmi, {\it {Duality in the Presence of Supersymmetry
  Breaking}},  {\em Science} (Jan., 1998) 27,
  [\href{http://xxx.lanl.gov/abs/9801146}{{\tt 9801146}}].

\bibitem{Arkani-Hamed1998a}
N.~Arkani-Hamed and R.~Rattazzi, {\it {Exact results for non-holomorphic masses
  in softly broken supersymmetric gauge theories}},  {\em Physics Letters B}
  {\bf 454} (May, 1999) 290--296, [\href{http://xxx.lanl.gov/abs/9804068}{{\tt
  9804068}}].

\bibitem{Kobayashi2000}
T.~Kobayashi and K.~Yoshioka, {\it {New RG-invariants of soft supersymmetry
  breaking parameters}},  {\em Physics Letters B} {\bf 486} (July, 2000)
  223--227, [\href{http://xxx.lanl.gov/abs/0004175}{{\tt 0004175}}].

\bibitem{Nelson2001}
A.~E. Nelson and M.~J. Strassler, {\it {Exact Results for Supersymmetric
  Renormalization and the Supersymmetric Flavor Problem}},  {\em Journal of
  High Energy Physics} {\bf 2002} (July, 2002) 021--021,
  [\href{http://xxx.lanl.gov/abs/0104051}{{\tt 0104051}}].

\bibitem{Luty1999}
M.~A. Luty and R.~Rattazzi, {\it {Soft supersymmetry breaking in deformed
  moduli spaces, conformal theories, and Script N = 2 Yang-Mills theory}},
  {\em Journal of High Energy Physics} {\bf 1999} (Nov., 1999) 1--37,
  [\href{http://xxx.lanl.gov/abs/9908085}{{\tt 9908085}}].

\bibitem{Abel2011}
S.~Abel, M.~Buican, and Z.~Komargodski, {\it {Mapping Anomalous Currents in
  Supersymmetric Dualities}},  {\em Physics} (May, 2011) 23,
  [\href{http://xxx.lanl.gov/abs/1105.2885}{{\tt arXiv:1105.2885}}].

\bibitem{Weinberg1998}
S.~Weinberg, {\it {Nonrenormalization Theorems in Nonrenormalizable Theories}},
   {\em Physical Review Letters} {\bf 80} (Apr., 1998) 3702--3705,
  [\href{http://xxx.lanl.gov/abs/9803099}{{\tt 9803099}}].

\bibitem{Jack1999}
I.~Jack and D.~Jones, {\it {Quasi-infrared fixed points and renormalization
  group invariant trajectories for nonholomorphic soft supersymmetry
  breaking}},  {\em Physical Review D} {\bf 61} (Mar., 2000) 095002,
  [\href{http://xxx.lanl.gov/abs/9909570}{{\tt 9909570}}].

\bibitem{Cecotti1986}
S.~Cecotti, L.~Girardello, and M.~Porrati, {\it {Constraints on partial
  super-Higgs}},  {\em Nuclear Physics B} (1986).

\bibitem{Cecotti1984}
S.~Cecotti, L.~Girardello, and M.~Porrati, {\it {Two into one won't go}},  {\em
  Physics Letters B} (1984).

\bibitem{Antoniadis1996}
I.~Antoniadis, H.~Partouche, and T.~R. Taylor, {\it {Spontaneous breaking of N=
  2 global supersymmetry}},  {\em Physics Letters B} {\bf 372} (1996),
  no.~April 83--87.

\bibitem{Antoniadis1996a}
I.~Antoniadis and T.~R. Taylor, {\it {Dual N = 2 SUSY Breaking}},  {\em
  Fortschritte der Physik/Progress of Physics} {\bf 44} (Apr., 1996) 487--492,
  [\href{http://xxx.lanl.gov/abs/9604062}{{\tt 9604062}}].

\bibitem{Partouche1997}
H.~Partouche and B.~Pioline, {\it {Partial spontaneous breaking of global
  supersymmetry}},  {\em Nuclear Physics B - Proceedings Supplements} {\bf 56}
  (July, 1997) 322--327, [\href{http://xxx.lanl.gov/abs/9702115}{{\tt
  9702115}}].

\bibitem{Seiberg1994b}
N.~Seiberg and E.~Witten, {\it {Monopoles, duality and chiral symmetry breaking
  in N = 2 supersymmetric QCD}},  {\em Nuclear Physics B} {\bf 431} (Dec.,
  1994) 484--550, [\href{http://xxx.lanl.gov/abs/9408099}{{\tt 9408099}}].

\bibitem{Seiberg1994c}
N.~Seiberg and E.~Witten, {\it {Electric-magnetic duality, monopole
  condensation, and confinement in N=2 supersymmetric Yang-Mills theory}},
  {\em Nuclear Physics B} {\bf 426} (Sept., 1994) 19--52,
  [\href{http://xxx.lanl.gov/abs/9407087}{{\tt 9407087}}].

\bibitem{Eguchi1995}
T.~Eguchi and S.-K. Yang, {\it {Prepotentials of N = 2 Supersymmetric Gauge
  Theories and Soliton Equations}},  {\em Modern Physics Letters A} {\bf 11}
  (Jan., 1996) 131--138, [\href{http://xxx.lanl.gov/abs/9510183}{{\tt
  9510183}}].

\bibitem{Komargodski2011}
Z.~Komargodski and A.~Schwimmer, {\it {On Renormalization Group Flows in Four
  Dimensions}},  \href{http://xxx.lanl.gov/abs/1107.3987}{{\tt
  arXiv:1107.3987}}.

\bibitem{Intriligator2004}
K.~Intriligator and B.~Wecht, {\it {RG fixed points and flows in SQCD with
  adjoints}},  {\em Nuclear Physics B} {\bf 677} (2004), no.~1-2 59.

\bibitem{Intriligator2003}
K.~Intriligator and B.~Wecht, {\it {The exact superconformal R-symmetry
  maximizes a}},  {\em Nuclear Physics B} {\bf 667} (Sept., 2003) 183--200,
  [\href{http://xxx.lanl.gov/abs/0304128}{{\tt 0304128}}].

\bibitem{Sohnius1978}
M.~Sohnius, {\it {Supersymmetry and central charges}},  {\em Nuclear Physics B}
  {\bf 138} (1978) 109--121.

\bibitem{Ohta1986}
N.~Ohta, H.~Sugata, and H.~Yamaguchi, {\it {N = 2 harmonic superspace with
  central charges and its application to self-interacting massive
  hypermultiplets}},  {\em Annals of Physics} {\bf 172} (Nov., 1986) 26--39.

\bibitem{Fayet1974}
P.~Fayet and J.~Iliopoulos, {\it {Spontaneously broken supergauge symmetries
  and Goldstone spinors}},  {\em Physics Letters B} {\bf 51} (1974), no.~5
  461--464.

\bibitem{Ivanov1997}
E.~A. Ivanov and B.~M. Zupnik, {\it {Modified N=2 supersymmetry and
  Fayet-Iliopoulos terms}},  \href{http://xxx.lanl.gov/abs/9710236}{{\tt
  9710236}}.

\bibitem{Fujiwara2006}
K.~Fujiwara, H.~Itoyama, and M.~Sakaguchi, {\it {Spontaneous Partial Breaking
  of N = 2 Supersymmetry and the U(N) Gauge Model}},  in {\em AIP Conference
  Proceedings}, vol.~903, pp.~521--524, AIP, Nov., 2007.
\newblock \href{http://xxx.lanl.gov/abs/0611284}{{\tt 0611284}}.

\bibitem{Fujiwara2006a}
K.~Fujiwara, {\it {Partial breaking of supersymmetry and decoupling limit of
  Nambu–Goldstone fermion in gauge model}},  {\em Nuclear Physics B} {\bf
  770} (May, 2007) 145--153, [\href{http://xxx.lanl.gov/abs/0609039}{{\tt
  0609039}}].

\bibitem{Fujiwara2005a}
K.~Fujiwara, H.~Itoyama, and M.~Sakaguchi, {\it {Partial breaking of
  supersymmetry and of gauge symmetry in the gauge model}},  {\em Nuclear
  Physics B} {\bf 723} (Sept., 2005) 33--52,
  [\href{http://xxx.lanl.gov/abs/0503113}{{\tt 0503113}}].

\bibitem{Fujiwara2004a}
K.~Fujiwara, H.~Itoyama, and M.~Sakaguchi, {\it {U(N) Gauge Model and Partial
  Breaking of N=2 Supersymmetry}},  \href{http://xxx.lanl.gov/abs/0410132}{{\tt
  0410132}}.

\bibitem{Fujiwara2004}
K.~Fujiwara, H.~Itoyama, and M.~Sakaguchi, {\it {Supersymmetric U ( N ) Gauge
  Model and Partial Breaking of N =2 Supersymmetry}},  {\em Progress of
  Theoretical Physics} {\bf 113} (Feb., 2005) 429--455,
  [\href{http://xxx.lanl.gov/abs/0409060}{{\tt 0409060}}].

\bibitem{Galperin2001}
A.~S. Galperin, E.~A. Ivanov, V.~I. Ogievetsky, and E.~S. Sokatchev, {\em
  {Harmonic Superspace}}.
\newblock Cambridge University Press, Cambridge, 1st~ed., 2001.

\bibitem{Galperin1985}
A.~S. Galperin, E.~A. Ivanov, S.~Kalitzin, V.~Ogievetsky, and E.~Sokatchev,
  {\it {Unconstrained N = 2 matter, Yang-Mills and supergravity theories in
  harmonic superspace}},  {\em Classical and Quantum Gravity} {\bf 2} (Jan.,
  1985) 127--127.

\bibitem{Fujiwara2005}
K.~Fujiwara, H.~Itoyama, and M.~Sakaguchi, {\it {Partial supersymmetry breaking
  and gauge model with hypermultiplets in harmonic superspace}},  {\em Nuclear
  Physics B} {\bf 740} (Apr., 2006) 58--78,
  [\href{http://xxx.lanl.gov/abs/0510255}{{\tt 0510255}}].

\bibitem{Itoyama2008}
H.~Itoyama, K.~Maruyoshi, and M.~Sakaguchi, {\it {N=2 quiver gauge model and
  partial supersymmetry breaking}},  {\em Nuclear Physics B} {\bf 794} (May,
  2008) 216--230, [\href{http://xxx.lanl.gov/abs/0709.3166}{{\tt
  arXiv:0709.3166}}].

\bibitem{Gates1984}
S.~J. Gates, {\it {Superspace formulation of new non-linear sigma models}},
  {\em Nuclear Physics B} {\bf 238} (1984) 349--366.

\bibitem{Fayet1976}
P.~Fayet, {\it {Fermi-Bose hypersymmetry}},  {\em Nuclear Physics B} {\bf 13}
  (1976), no.~1 135--155.

\bibitem{Salam1975}
A.~Salam and J.~Strathdee, {\it {Superfields and fermi-bose symmetry}},  {\em
  Physical Review D} {\bf 11} (1975), no.~6 1521.

\bibitem{Argyres1996}
P.~C. Argyres, M.~{Ronen Plesser}, and N.~Seiberg, {\it {The moduli space of
  vacua of N = 2 SUSY QCD and duality in N = 1 SUSY QCD}},  {\em Nuclear
  Physics B} {\bf 471} (July, 1996) 159--194,
  [\href{http://xxx.lanl.gov/abs/9603042}{{\tt 9603042}}].

\bibitem{Fayet1986}
P.~Fayet, {\it {Six-dimensional supersymmetric QED, R-invariance and N = 2
  supersymmetry breaking by dimensional reduction}},  {\em Nuclear Physics B}
  {\bf 263} (Jan., 1986) 649--686.

\bibitem{Seiberg1988}
N.~Seiberg, {\it {Supersymmetry and non-perturbative beta functions}},  {\em
  Physics Letters B} {\bf 206} (1988), no.~1 75--80.

\bibitem{Zupnik1987}
B.~M. Zupnik, {\it {The action of the supersymmetric N= 2 gauge theory in
  harmonic superspace}},  {\em Physics Letters B} {\bf 183} (1987), no.~2
  175--176.

\bibitem{Leigh1995}
R.~G. Leigh and M.~J. Strassler, {\it {Exactly marginal operators and duality
  in four dimensional N = 1 supersymmetric gauge theory}},  {\em Nuclear
  Physics B} {\bf 447} (July, 1995) 95--133,
  [\href{http://xxx.lanl.gov/abs/9503121}{{\tt 9503121}}].

\bibitem{Strassler1996}
M.~J. Strassler, {\it {Manifolds of Fixed Points and Duality in Supersymmetric
  Gauge Theories}},  {\em Progress of Theoretical Physics Supplement} {\bf 123}
  (Feb., 1996) 373--380, [\href{http://xxx.lanl.gov/abs/9602021}{{\tt
  9602021}}].

\end{thebibliography}\endgroup
\end{document}